\documentclass[]{aa} 

\usepackage{graphicx}
\usepackage{txfonts}
\usepackage{xcolor}
\usepackage{tabularx}
\usepackage{chemformula}

\newcommand{\waspT}{$T_{\textrm{eq}}=2358\pm 52$ K;}
\newcommand{\wasphotT}{$T_{\textrm{eq}}=2641 \pm 34$ K;}
\newcommand{\MatwoT}{$T_{\textrm{eq}}\sim 2240 $ K;}
\newcommand{\MafiveT}{$T_{\textrm{eq}} = 2181 $ K;}
\usepackage[pdfpagelabels=false]{hyperref}      
\hypersetup{colorlinks=true,linkcolor=blue,citecolor=blue,filecolor=blue,urlcolor=blue,}

\def\figurenum#1{\def\thefigure{#1}\let\currentlabel\thefigure}

\begin{document} 

   \title{The Mantis network.\\ {I}. A standard grid of templates and masks for cross-correlation analyses of ultra-hot Jupiter transmission spectra\thanks{ Templates and line masks are available at the CDS via anonymous
     ftp to \url{https://cdsarc.cds.unistra.fr/viz-bin/cat/J/A+A/669/A113} (\url{ftp://130.79.128.5}) or via \url{https://cdsarc.cds.unistra.fr/viz-bin/cat/J/A+A/669/A113}}}
   \titlerunning{Mantis. I. A standard grid of templates and masks}
   \authorrunning{Kitzmann, Hoeijmakers, Grimm et al.}

   \author{D. Kitzmann\inst{1}
          \and
          H. J. Hoeijmakers\inst{1,2}
          \and 
          S. L. Grimm\inst{1}
          \and
          N. W. Borsato\inst{2}
          \and
          A. Lueber\inst{1}
          \and
          B. Prinoth\inst{2}
          }

   \institute{University of Bern, Center for Space and Habitability, Gesellschaftsstrasse 6, CH-3012, Bern, Switzerland\\
              \email{daniel.kitzmann@unibe.ch, simon.grimm@unibe.ch}
              \and
              Lund Observatory, Department of Astronomy and Theoretical Physics, Lund University, Box 43, 221 00 Lund, Sweden\\
              \email{jens.hoeijmakers@astro.lu.se}
             }

   \date{Received ; accepted}

  \abstract{The atmospheres of ultra-hot Jupiters are highly interesting and unique chemical laboratories. Due to the very high atmospheric temperatures, their chemical composition is dominated by atoms and ions instead of molecules, and the formation of aerosols on their day-sides is unlikely. 
  Thus, for these planets detailed chemical characterisations via the direct detection of elements through high-resolution day-side and transit spectroscopy are possible. This in principle allows the element abundances of these objects to be directly inferred, which may provide crucial constraints on their formation process and evolution history.  
  
  In the recent past, several chemical species, mostly in the form of atoms and ions, have already been detected using high-resolution spectroscopy in combination with the cross-correlation technique.  \\
  As part of the Mantis network, we provide a grid of standard templates in this study, designed to be used together with the cross-correlation method. This allows for the straightforward detection of chemical species in the atmospheres of hot extrasolar planets.
  In total, we calculate high-resolution templates for more than 140 different species across several atmospheric temperatures. In addition to the high-resolution templates, we also provide line masks that just include the position of line peaks and their absorption depths relative to the spectral continuum. A separate version of these line masks also takes potential blending effects with lines of other species into account. All templates and line masks are publicly available on the CDS data server.}
  

  \maketitle

\section{Introduction}

High-resolution ground-based spectroscopy of exoplanet atmospheres has rapidly developed since the inception of cross-correlation-based analysis methods \citep{Brown2001,Snellen2010}. These methods rely on averaging the absorbed or emitted flux of a known ensemble of spectral lines to measure their collective presence and strength, even though individual lines may be significantly weaker than the uncertainty (signal-to-noise ratio) at any particular wavelength. This method was initially applied at infrared wavelengths, where absorption bands of H$_2$O and CO in particular provide large numbers of absorption lines that can be combined via cross-correlation.

High-resolution spectroscopy is uniquely sensitive to the dynamics of the planet and its atmosphere, because such instruments have the ability to resolve radial velocities of the order of 1 km s$^{-1}$. Additionally, this facilitates direct measurements of the orbital velocity of exoplanets and thus resolves degeneracies in the orbital inclination  or the stellar mass \citep{Snellen2010,Brogi2012}. On the other hand, the large day-to-night temperature contrasts exhibited by many hot-Jupiter-type exoplanets may elicit vigorous atmospheric dynamics, commonly resulting in wind speeds of several km s$^{-1}$. This may result in detectable Doppler shifts of the observed planetary absorption spectrum, or Doppler-broadening in excess of the rotation velocity caused by the planet's diurnal spin.

Gas giant exoplanets with sufficiently high temperatures feature rich transmission spectra at optical wavelengths due to the presence of atomic and ionic metals \citep{Kitzmann2018ApJ...863..183K, Lothringer2018ApJ...866...27L}. The manifestation of absorption signatures in transmission spectra depends on the temperature structure of the atmosphere and various chemical processes (including condensation, dissociation, and ionisation; e.g. \citealt{HK2017}). Absorption signals of a multitude of atoms and ions therefore provide powerful probes of the thermochemical nature of these extreme atmospheres, especially in concert with absolutely calibrated low-resolution spectrographs aboard space observatories \citep{Brogi2017ApJ...839L...2B}.

Directly detecting atoms and ions in these hot atmospheres and constraining their abundances would in principle allow the corresponding element abundances to be derived. In atmospheres of cooler planets, deriving metallicities is much more complicated because elements are usually bound in a multitude of molecules or even condensates.

\begin{table*}
  \caption[]{Overview of existing and planned high-resolution spectrographs.}
  \label{tab:high-res_spectrographs}
  \centering
  \begin{tabularx}{0.88\textwidth}{lcrccc}
  \hline\hline
  Spectrograph & Wavelength coverage (nm) & Resolution & Observatory & Mirror Diameter (m) &  References \\
  \hline
  HARPS    & 383 -- 693   & 120\,000  & La Silla     & 3.6          & 1\\
  NIRPS    & 950 -- 1800  & 100\,000  & La Silla     & 3.6          & 25, * \\
  HARPS-N  & 383 -- 690   & 115\,000  & TNG          & 3.6          & 2\\
  GIANO    & 950 -- 2450  & 50\,000   & TNG          &              & 8 \\
  ESPRESSO & 380 -- 788   & 70\,000   & VLT          & 8.2 -- 16    & 3 \\
           &              & 140\,000  &              &              &   \\
           &              & 190\,000  &              &              &  \\
  CRIRES+  & 950 -- 1120  & 100\,000  & VLT          & 8.2          & 4, 5 \\
           & 1116 -- 1362 &           &              &              &   \\
           & 1423 -- 1769 &           &              &              &   \\
           & 1972 -- 2624 &           &              &              &   \\
           & 2869 -- 4188 &           &              &              &   \\
           & 3583 -- 5300 &           &              &              &   \\
  UVES     & 300 -- 500   & 80\,000   & VLT          & 8.2          & 6 \\
           & 420 -- 1100  & 110\,000  &              &              &   \\
  CARMENES & 520 -- 960   & 94\,600   & Calar Alto   & 3.5          & 7 \\
           & 960 -- 1710  & 80\,400   &              &              &   \\
  PEPSI    & 384 -- 913   & 50\,000   & LBT          & 2$\times$8.4 & 9, 10 \\
           &              & 130\,000  &              &              &   \\
           &              & 250\,000  &              &              &   \\
  HIRES    & 300 -- 1100  & 67\,000   & Keck         & 10           & 11 \\
  NIRSPEC  & 950 -- 5500  & 25\,000   & Keck         & 10           & 12\\
  MAROON-X & 500 -- 920   & 80\,000   & Gemini North & 8            & 13\\
  GRACES   & 400 -- 1000  & 66\,000   & Gemini North & 8.1          & 14 \\
  ESPaDonS & 370 -- 1050  & 81\,000   & CFHT         & 3.6          & 15 \\
  SPIRou   & 950 -- 2500  & 71\,000   & CFHT         &              & 16 \\
  IGRINS   & 1450 -- 2450 & 45\,000   & Gemini South & 8            & 17 \\
  HDS      & 300 -- 1000  &  90\,000  & Subaru       & 8.2          & 18 \\
           &              & 160\,000  &              &              &    \\
  IRD      &  970 -- 1750 &  70\,000  & Subaru       & 8.2          & 19 \\
  FIES     & 370 -- 730   &  25\,000  & NOT          & 2.6          & 20 \\
           &              &  46\,000  &              &              &   \\
           &              &  67\,000  &              &              &   \\
 HARPS3    & 380 -- 690   & 115\,000  & INT          & 2.5          & 21 \\
 Veloce    & 580 -- 930   & 75\,000   & AAT          & 3.9          & 22 \\
 EXPRES    & 400 -- 680   & 137\,500  & LDT          & 4.3          & 23 \\
 NEID      & 380 -- 930   &  60\,000  & WIYN         & 3.5          & 24 \\
           &              &  90\,000  &              &              &   \\
  HIRES    & 550 -- 1800  & 100\,000  & ELT          & 39           & 26, * \\
  \hline
  \end{tabularx}
  \tablebib{
    (1) \citet{MayorHARPS2003Msngr.114...20M};
    (2) \citet{CosentinoHARPSN2012SPIE.8446E..1VC};
    (3) \citet{PepeESPRESSO2021A&A...645A..96P};
    (4) \citet{KaeuflCRIRES2004SPIE.5492.1218K};
    (5) \citet{DornCRIRES2014Msngr.156....7D};
    (6) \citet{DekkerUVES000SPIE.4008..534D};
    (7) \citet{QuirrenbachCARMENES2016SPIE.9908E..12Q};
    (8) \citet{GIANO2012SPIE.8446E..3TO};
    (9) \citet{StrassmeierPEPSI2018SPIE10702E..12S};
    (10) \citet{StrassmeierPEPSI2015AN....336..324S};
    (11) \citet{VogtHIRES1994SPIE.2198..362V};
    (12) \citet{McLeanNIRSPEC1998SPIE.3354..566M};
    (13) \citet{Seifahrt2018SPIE10702E..6DS};
    (14) \citet{Chene2014};
    (15) \citet{Donati2006};
    (16) \citet{Donati2020};
    (17) \citet{Park2014SPIE.9147E..1DP};
    (18) \citet{Noguchi2002};
    (19) \citet{Tamura2012};
    (20) \citet{Telting2014};
    (21) \citet{Thompson2016};
    (22) \citet{Gilbert2018};
    (23) \citet{Petersburg2020};
    (24) \citet{Schwab2016};
    (25) \citet{Wildi2017SPIE10400E..18W};
    (26) \citet{Marconi2021Msngr.182...27M};
    * These instruments are still under development, and their characteristics are subject to change.
  }
\end{table*}

The discovery of iron and titanium in the transmission spectrum of KELT-9\,b \citep{Hoeijmakers2018Natur.560..453H} has motivated numerous groups to carry out studies using cross-correlation-based analyses to attempt to detect and measure signatures of metals. These efforts have led to the detection of a plurality of metals in a growing number of planets with temperatures over $\sim 2000$ K \citep[e.g.][a.o.]{Hoeijmakers2018Natur.560..453H,Hoeijmakers2019A&A...627A.165H, borsa2021atmospheric, tabernero2021espresso}. However, the recent literature contains a variety of methods and practices, and analyses of the same planet or observation are not guaranteed to yield consistent results. For example, \citet{Ben-Yami2020} reported a detection of Fe$^+$ absorption in WASP-121\,b \citep[\waspT][]{Delrez2016}, while \citet{Hoeijmakers2020} reported a non-detection based on the same data, recently reinforced by \citet{Merritt2021}. Some results are poorly understood, such as the strong detection of metal ions in MASCARA-2\,b \citep[\MatwoT][]{Casasayas-Barris2020MASCARA,Hoeijmakers2020MASCARA}, which is significantly cooler than WASP-121\,b, or an apparent lack of metal absorption in the transmission spectrum of MASCARA-5\,b \citep[\MafiveT][]{Stangret2021}, while the spectrum of the similar WASP-189\,b \citep[\wasphotT][]{Anderson2018} is rich in absorption lines of metals \citep{Prinoth2022}. Emission by the TiO molecule was reported in the atmosphere of UHJ WASP-33\,b by \citet{Nugroho2017high}, but subsequent analysis has failed to confirm this result \citep{herman2020search,Serindag2021A&A...645A..90S,Espinoza2019MNRAS.482.2065E,Sedaghati2021MNRAS.505..435S}. \\

Integral to the application of cross-correlation-based techniques is the use of a template spectrum to combine spectral lines of interest. If the objective is to detect a species at high confidence or when the cross-correlation operation is used to match the model spectrum to the data, a template is designed to most closely match the true transmission spectrum, which is usually poorly known a priori. Computing accurate models at high resolution involves challenges related to line-list accuracy, computational resources, and theoretical assumptions. Some studies use full numerical simulations to compute cross-correlation templates, thereby introducing particular model dependences that may vary from case to case. Others use versatile analytical approximations \citep[e.g.][]{Merritt2021}, while yet others make use of cross-correlation functions generated by the data-reduction pipelines of high-precision radial velocity instruments such as HARPS and ESPRESSO, which are created using stellar masks that contain a number of absorbing species simultaneously \citep{Bourrier2020,Ehrenreich2020}. 

Furthermore, various different implementations of the cross-correlation function exist. Many studies make use of cross-correlation coefficients that are evaluated by multiplying observed and model spectra that are defined on the same wavelength grid \citep[e.g.][]{Snellen2010,Brogi2012,Birkby2017,Hoeijmakers2018Natur.560..453H,Ehrenreich2020,Casasayas-Barris2020MASCARA,pino2020neutral,Gibson2020}. Common approaches are: the Pearson correlation coefficient, which is a normalised, unit-less quantity that expresses the degree of correlation or anti-correlation; the simple weighted average that preserves the unit of the transmission spectrum (i.e. normalised transit depth); and variations that cast these quantities into measures of the statistical likelihood, which are maximised to enable model parameter fitting \citep[][a.o.]{BrogiLine2019,Gibson2020}. Alternatively, another form of cross-correlation directly uses lists of line positions, by which individual samples of the observed spectra are weighted directly. These are sometimes termed binary masks\footnote{Binary mask implies that equal weight is applied to the selected spectral lines, though this is not necessarily needed.} and treat spectral lines as discrete approximations to the infinitely narrow Dirac-$\delta$ function. Such masks are commonly applied to precision radial-velocity measurements in attempts to discover exoplanets \citep{Baranne1996} as well as to detect species in exoplanet atmospheres \citep[e.g.][]{Allart2017}. 

This plurality of practices adds a layer of complexity to the cross-comparison of studies of the transmission spectra of ultra-hot Jupiters and hinders the interpretation of apparent differences in results. This motivates efforts to standardise methodology. In this paper, we therefore present the calculation of a grid of high-resolution model transmission spectra at optical and near-infrared wavelengths, from which we derive cross-correlation templates and masks. This work is part of the Mantis network, an international research collaboration dedicated to high-resolution spectroscopy. We make these publicly available with the intention to improve the comparability of cross-correlation methods and derived results and to benefit users who have limited access to the expertise needed to produce templates or who wish to benchmark their own templates.

\section{Overview of high-resolution spectrographs}

Several high-resolution spectrographs are already available on a wide range of different telescopes. Table \ref{tab:high-res_spectrographs} provides an overview of existing and planned instruments, along with the observatories they are located at and the corresponding telescopes' mirror diameters.

These high-resolution spectrographs offer a wide range of different wavelength coverages and spectral resolutions. The resolutions can vary from 25\,000 for NIRSPEC up to 250\,000 for PEPSI, while the spectral coverage can range from 300 nm for UVES up to 5.5 $\mu$m for the NIRSPEC spectrograph at the Keck Observatory. In addition to the different instrument properties, also the mirror sizes differ significantly between 3.5 metres and 39 metres for the various different telescopes. 

The templates provided in this study are designed to be used with any of these spectrographs. The smallest spectral resolution in our templates is at least twice as high as the highest spectral resolution offered by PEPSI, while the maximum resolution is by more than a factor of ten higher (see Sect. \ref{sec:spectra_calculations} and Fig. \ref{fig:spectral_resolution}). In terms of wavelengths, our templates provide a range from 0.32 $\mu$m to 5.5 $\mu$m. This range is chosen because the templates are mainly created to correlate for the lines of atoms and ions in spectra of hot Jupiter atmospheres. These lines are usually located in the visible and near-infrared wavelength regions. The chosen wavelength range covers all high-resolution spectrographs listed in Table \ref{tab:high-res_spectrographs}.

\section{Cross-correlation templates}
Equation \ref{eq:ccf} gives a comparison of the most common implementations of the cross-correlation operation applied to a time series of spectra:

\begin{equation}\label{eq:ccf}
    C(v,t) = \begin{cases}
    \frac{\sum x(t)T(v)}{\sum T(v)} & \textrm{weighted average} \\ 
    \frac{\sum \tilde{x}(t) \tilde{T}(v)}{\sqrt{ \sum \tilde{x}(t)^2 \sum \tilde{T}(v)^2 }} & \textrm{Pearson Correlation}
    \end{cases} \ , 
\end{equation}where $\sum$ denotes summation over all spectral samples that make up the high-resolution spectrum $x(t)$ taken at time $t$ (meaning that $x$ is a vector of flux values recorded by a spectrograph). The $T(v)$ is the cross-correlation template Doppler-shifted to a radial velocity, $v$, generating a cross-correlation coefficient, $C(v,t)$, for a range of velocity shifts and for each of the spectra $x(t)$ in the time series. In the case of the Pearson correlation, $\tilde{x}$ and $\tilde{T}$ are the spectra and the template minus their averages: $\tilde{x} = x(t)-\overline{x}(t)$ and  $\tilde{T} = T(v)-\overline{T}(v)$.\\

The vector $x$ is composed of a large number of $N$ flux values, which are accompanied by $N$ corresponding values of wavelength $\lambda$. In principle, the cross-correlation is agnostic to the fact that values of $\lambda$ are associated with $x$, or that $x$ carries the physical meaning of a spectrum. However, to carry out the multiplication of the data, $x$, with the template, $T$, the template needs to be evaluated onto the same wavelength grid as $x$.
Many studies in the literature that employ the cross-correlation method use models of the expected exoplanet spectrum to act as templates. These models may be evaluated at much higher spectral resolution than the observed spectrum $x$, which means that a template $T$ needs to be interpolated onto $\lambda$, for each value of the Doppler velocity $v$. In order to include only values of $\lambda$ that are inside spectral lines of interest and not in the continuum, the template must be continuum-subtracted, such that the continuum between the spectral lines is zero (or negligible, as the wings of spectral lines diminish with the distance to their line centres).\\

Instead of being a vector of $N$ values that matches the wavelength positions of $x$, a template may also be defined as a list of singular line positions and relative strengths, sometimes referred to as a binary mask. In this case, $T(v)$ is evaluated by distributing the weight of each line over the two spectral samples surrounding the central wavelength of the line. This means that $T(v)=0$ at any $\lambda$ away from any spectral line. As such, the evaluation of the cross-correlation simplifies to include only summations over values of $x$ where $T$ is non-zero. Thus, the summation is performed over at most two times the number of spectral lines included in the line list that are in the wavelength region covered by the spectrograph. However, regardless of how the cross-correlation operation is implemented, the weight assigned to each spectral line is prescribed by a physical expectation of the contents of the target spectrum. Therefore, we start the creation of cross-correlation templates with a model of the exoplanet atmosphere.

\subsection{General atmosphere description}

We assumed a generic hot Jupiter atmosphere, using a representative surface gravity of $g=2000 \ \mathrm{cm \, s^{-2}}$ and a planetary radius $R_p$ of 1.5 Jupiter radii at a pressure of 10 bar. Because of the normalisation of the template over the wavelength range, the cross-correlation template is only sensitive to relative line depths, which are largely insensitive to the choice in $g$ and $R_p$ in the limit that the scale height $H$ is small compared to $R_p$. The atmosphere is extended to a pressure of $10^{-15} \ \rm{bar}$ to allow the cores of even the strongest spectral lines to be fully resolved in the transmission spectra. For simplicity, we assumed a constant isothermal temperature profile throughout the atmosphere, with the temperature $T$ as a free parameter. 

Following our previous approaches in, for example, \citet{Kitzmann2018ApJ...863..183K} or \citet{Hoeijmakers2019A&A...627A.165H}, we prescribed the chemical composition of the atmosphere by assuming equilibrium chemistry, which is likely to be a good approximation for most species and pressures, considering the high equilibrium temperatures of ultra-hot Jupiters. We used the open-source chemistry code \textsc{FastChem 2.0}\footnote{\url{https://github.com/exoclime/FastChem}} \citep{Stock2018MNRAS.479..865S, Stock2022MNRAS.tmp.2449S} to calculate the mixing ratios of all species as a function of temperature and pressure. We limited the determination of the chemical composition to solar elemental abundances. The standard \textsc{FastChem} release only contains molecular species and ions for elements at least as abundant as germanium. Therefore, mass action constants for all the atomic and ionic species up to uranium not included in the standard release of \textsc{FastChem} were added in \citet{Hoeijmakers2019A&A...627A.165H} and are listed in their appendix. 

For the chemistry calculations in this study, we now also added additional molecules for all elements up to uranium. In particular, we added all molecules from the \citet{Barin1997thermochemical} database that were not already part of the standard \textsc{FastChem} release. In total, we now include almost 1100 molecules and ions for 81 elements. 

With the pressure-dependent mean molecular weight provided by \textsc{FastChem 2.0}, we finally converted the atmospheric pressure to geometrical altitudes assuming hydrostatic equilibrium. The surface gravity is thereby allowed to decrease with altitude.

\begin{table*}
  \caption[]{Opacity species and references used in the calculations of the transmission spectra.}
  \label{tab:opacity_species}
  \centering
  \begin{tabularx}{0.98\textwidth}{llXl}
  \hline\hline
  Linelist & Type & Species &  References \\
  \hline
         & Rayleigh scattering  & \ch{H2}                   & 1\\
         &                      &H                          & 2\\
         &                      &He                         & 3, 4\\
         &                      &                           & \\
         & Continuum absorption & \ch{H-} (free-free \& bound-free) & 5\\
         &                      & \ch{H2-} (free-free)      & 6\\
                                &                           & \\
  HITRAN & CIA                  & \ch{H2}-\ch{H2}           & 7\\
         &                      & \ch{H2}-He                & 8\\
         &                      & H-He                      & \\
         &                      &                           & \\
  VALD3  & Line absorption      &Li, Be, B, N, O, F, Ne, Na, Al, Si, P, S, Cl, K, Ca, Sc, Ti, V, Cr, Mn, Fe, Co, Ni, Cu, Zn, Ga, Ge, As, Se, Rb, Sr, Y, Zr, Nb, Mo, Ru, Rh, Pd, Ag, Cd, In, Sn, Sb, Te, Cs, Ba, La, Ce, Pr, Nd, Sm, Eu, Gd, Tb, Dy, Ho, Er, Tm, Yb, Lu, Hf, Ta, W, Re, Os, Ir, Pt, Au, Hg, Tl, Pb, Bi, Th, U, & 9, 10\\
         &                      &Li$^+$, Be$^+$, B$^+$, N$^+$, O$^+$, F$^+$, Ne$^+$, Na$^+$, Mg$^+$, Al$^+$, Si$^+$, P$^+$, S$^+$, Ca$^+$, Cl$^+$, Ar$^+$, K$^+$, Ca$^+$, Sc$^+$, Ti$^+$, V$^+$, Cr$^+$, Mn$^+$, Fe$^+$, Co$^+$, Ni$^+$, Cu$^+$, Zn$^+$ Ga$^+$, Ge$^+$, Y$^+$, Zr$^+$, Nb$^+$, Mo$^+$, Ru$^+$, Rh$^+$, Pd$^+$, Ag$^+$, Cd$^+$, In$^+$, Sn$^+$, Xe$^+$, Ba$^+$, La$^+$, Ce$^+$, Pr$^+$, Nd$^+$, Sm$^+$, Eu$^+$, Gd$^+$, Tb$^+$, Dy$^+$, Ho$^+$, Er$^+$, Tm$^+$, Yb$^+$, Lu$^+$, Hf$^+$, Ta$^+$, W$^+$, Re$^+$, Os$^+$, Ir$^+$, Pt$^+$, Au$^+$, Hg$^+$, Pb$^+$, Bi$^+$, Th$^+$, U$^+$ & \\
         &                      &                           & \\
  Kurucz & Line absorption      &H, He, C, Mg, Ar, He$^+$, Sr$^+$   & 11 \\
         &                      &                           & \\
  Exomol & Line absorption      &\ch{H2O}                   & 12\\
         &                      &CO                         & 13\\
         &                      &VO                         & 14\\
         &                      &TiO                        & 15\\
         &                      &FeH                        & 16\\
         &                      &OH                         & 17\\
  \hline
  \end{tabularx}
  \tablebib{
    (1) \citet{Cox2000asqu.book.....C};
    (2) \citet{Lee2004MNRAS.347..802L}; 
    (3) \citet{Sneep2005JQSRT..92..293S};
    (4) \citet{Thalman2014JQSRT.147..171T};
    (5) \citet{John1988A&A...193..189J};
    (6) \citet{Bell1980JPhB...13.1859B};
    (7) \citet{Abel_doi:10.1021/jp109441f};
    (8) \citet{Abel2012JChPh.136d4319A};
    (9) \citet{Ryabchikova2015PhyS...90e4005R}; 
    (10) \citet{Pakhomov2019ARep...63.1010P};
    (11) \citet{Kurucz2017CaJPh..95..825K};
    (12) \citet{Polyansky2018MNRAS.480.2597P};
    (13) \citet{Li2015ApJS..216...15L};
    (14) \citet{McKemmish2016MNRAS.463..771M};
    (15) \citet{McKemmish2019MNRAS.488.2836M};
    (16) \citet{Bernath2020JQSRT.24006687B};
    (17) \citet{Brooke2016JQSRT.168..142B}
  }
\end{table*}

\subsection{Opacity calculations}

For the calculation of the transmission spectra, we used absorption and scattering coefficients of various atomic and molecular species. A complete list of all species and the corresponding references to their corresponding sources is given in Table \ref{tab:opacity_species}.

Absorption cross-sections were generated for various atoms and ions found in the VALD3 and Kurucz database. We included all atoms and singly ionised species for which spectral and chemistry data are available. Higher ionised species are neglected here because their abundances and hence their spectral signatures are expected to be minor and buried in stronger bands of neutrals and single ions in the near-UV \citep{Hoeijmakers2019A&A...627A.165H}. 

The opacities were calculated using our \textsc{HELIOS-K} opacity calculator\footnote{\url{https://github.com/exoclime/HELIOS-K}} \citep{Grimm2015ApJ...808..182G, Grimm2021ApJS..253...30G}. Because we expect the cores of the atomic and ionic lines to originate from very low pressures in the transmission spectra, we did not use the available van der Waals or Stark broadening parameters. Instead, we only employed temperature-dependent Doppler broadening, as well as the natural line width (radiation dampening) in the calculation of the line opacities. An overview of these opacities can be found in the appendix of \citet{Grimm2021ApJS..253...30G}.

In addition to atoms and ions, we also included water (\ch{H2O}), carbon monoxide (CO), vanadium oxide (VO), titanium oxide (TiO), iron hydride (FeH), and the hydroxyl radical (OH), which are expected to absorb at higher pressures compared to atoms at optical wavelengths. Therefore, besides Doppler broadening and radiation dampening, we also used the pressure broadening coefficients from the Exomol database to describe the line shapes. A full description of the calculation of the opacities can be found in \citet{Grimm2021ApJS..253...30G}. We note that these opacity calculations assume local thermodynamic equilibrium, in particular that the species' levels are populated according to the Boltzmann statistics. Thus, spectral lines that form outside of this regime, and therefore depend strongly on the specific physical conditions of the atmospheres they are forming in, are not covered by the opacity calculations. This includes, for example, the metastable line of helium at 1083 nm that can form in escaping atmospheres of exoplanets \citep{Oklopcic2018ApJ...855L..11O}.

Important continuum absorption sources were added for collision-induced absorption (CIA) of \ch{H2-H2}, \ch{H2-He}, and \ch{H-He} collisions, as well as the free-free and bound-free absorption of \ch{H-} and \ch{H2-}- Especially \ch{H-} will dominate the overall continuum of the transmission spectra at higher temperatures. Finally, Rayleigh scattering cross-sections were added for the main atmospheric constituents: \ch{H2}, \ch{H}, and He.

\subsection{Transmission spectra calculations}
\label{sec:spectra_calculations}

To obtain the templates for the individual species, we first constructed transmission spectra for the assumed hot Jupiter atmosphere as a function of the isothermal temperature $T$. We cover a range of representative temperatures, starting at high temperatures where \ch{H-} is the main continuum absorber and down to temperatures, at which \ch{H2O} and \ch{CO} start to appear again. In total, we use five different temperatures: 5000 K, 4000 K, 3000 K, 2500 K, and 2000 K.

To generate the spectra, we used our observation simulator \textsc{Helios-o}, which has previously been used in, for example, \citet{Gaidos2017MNRAS.468.3418G} and \citet{Bower2019A&A...631A.103B}.
We calculated the apparent, wavelength-dependent radius of the planet with
\begin{equation}
    \pi R_{p,\lambda}^2 = \pi \left[R_{p}(z=0) + z_{\mathrm{eff}, \lambda} \right]^2 = \int_0^\infty 2 \pi r \left(1 - e^{-\tau_\lambda(r)} \right) \mathrm d r \ ,
    \label{eq:transit_radius}
\end{equation}
where $z$ is the vertical altitude coordinate, $R_{p}(z=0)$ the planetary radius at the bottom of the computational domain, and $z_{\mathrm{eff}}$ the effective tangent height. For the transmission spectra calculated in this study, we put the bottom of the computed atmosphere ($z=0$) at an atmospheric pressure of 10 bar and at a radius of $R_{p}(z=0) = 1.5 \, R_\mathrm{J}$.

The variable $\tau_\lambda(r)$ in Eq. \eqref{eq:transit_radius} is the slant optical depth for an impact parameter at a radius $r$. With $x$ as the spatial coordinate along that path, the optical depth is formally given by
\begin{equation}
    \tau_\lambda(r) = \int_{-\infty}^{+\infty} \chi_\lambda(x(r)) \, \mathrm d x \ ,
\end{equation}
where $\chi_\lambda(x(r))$ is the extinction coefficient (i.e. the sum of the scattering and absorption coefficients) along $x$.

The transmission spectra were calculated within the spectral range from 31\,250 $\mathrm{cm^{-1}}$ ($0.32 \ \mathrm{\mu m}$) to 1810 $\mathrm{cm^{-1}}$ ($\approx 5.525 \ \mathrm{\mu m}$) with a constant wavenumber step of $0.01 \ \rm{cm^{-1}}$. Each spectrum thus contains almost 3\,000\,000 data points. Due to the usage of a constant step size in wavenumber space, the spectral resolution $R = \lambda/\Delta \lambda$ is not constant over the entire spectral range.

The resulting spectral resolution as a function of wavelength varies from a value of roughly 3\,000\,000 in the visible wavelength region to about 183\,000 in the infrared (see also Fig. \ref{fig:spectral_resolution}). This is high compared to the resolving power of most existing high-resolution spectrographs for which these templates are intended, as well as upcoming instruments, in particular those planned for next-generation telescopes (see Table \ref{tab:high-res_spectrographs}).

\subsection{Single-species templates}
\label{sec:singe_species_templates}

Cross-correlation analyses are often performed using template spectra that contain lines of individual species, to isolate their contribution from the planet's transmission spectrum one species at a time. For this purpose, we calculated transmission spectra that include continuum-forming species and a single, line-forming template species in each case. These continuum species are those that provide a smooth Rayleigh scattering opacity, continuum absorption, or CIA species (see Table \ref{tab:opacity_species}). The template spectra were calculated for each line-forming species listed in Table \ref{tab:opacity_species} and for the five temperatures listed in the previous subsection. For each temperature, we also additionally provide a transmission spectrum that contains only the continuum-forming species. These can be used to remove the underlying continuum from the calculated single-species template spectra. In total, 750 of these template spectra were generated, including all the atoms, ions, and the six molecules listed in Table \ref{tab:opacity_species}.

\begin{figure}
  \resizebox{\hsize}{!}{\includegraphics{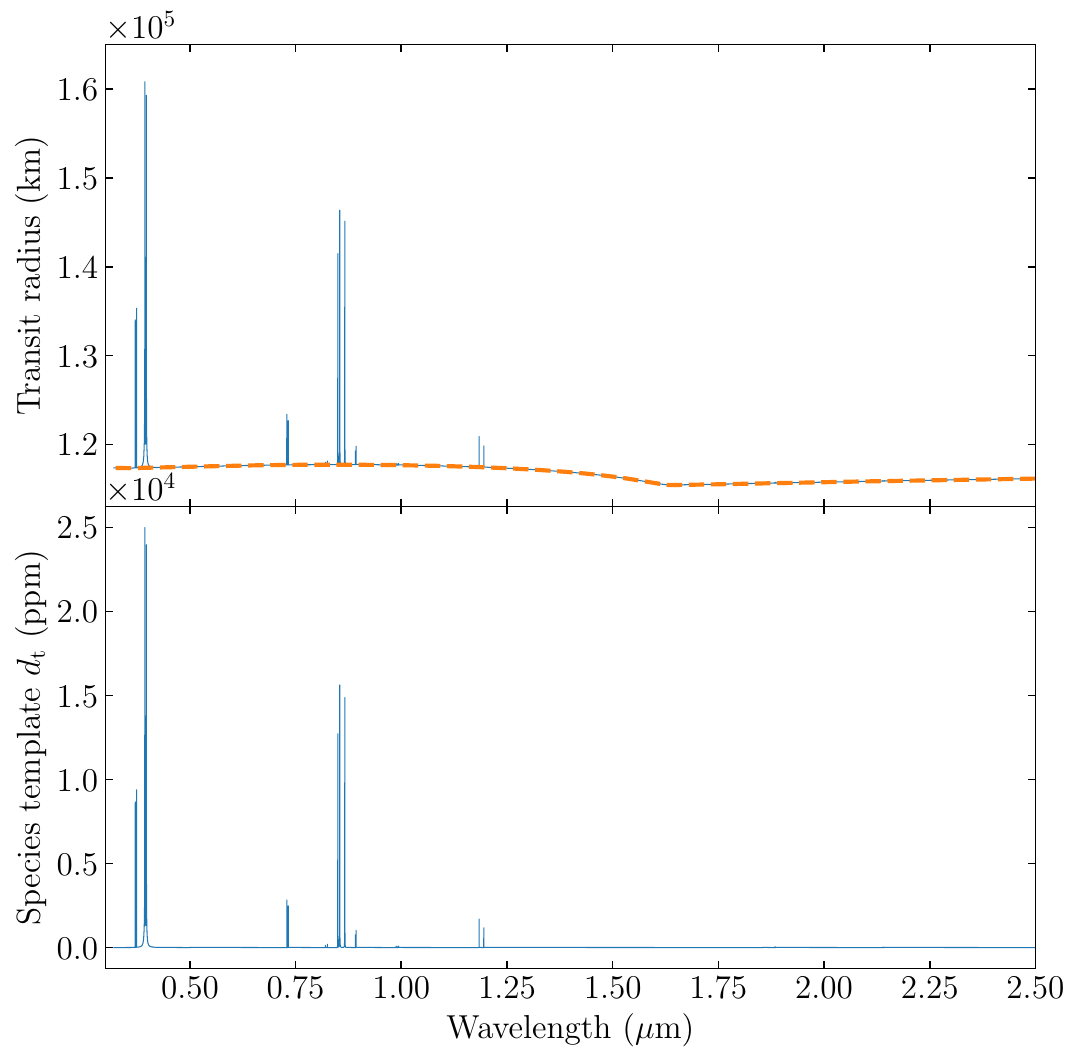}}
  \caption{Example of a template construction of \ch{Ca+} at an atmospheric temperature of $4000 \ \rm{K}$. The upper panel shows the transmission spectrum with \ch{Ca+} and additional continuum-opacity species (blue line), as well as one only containing the continuum-forming species (dashed orange line). The lower panel depicts the final template, created by subtracting the continuum and the \ch{Ca+} spectra from the top panel.}
  \label{fig:spectra_examples1}
\end{figure}

We present the templates for a species, $i$, in the form of the transit depth above the spectral continuum $d_{\mathrm{t},i}(\lambda)$, given by
\begin{equation}
    d_{\mathrm{t},i}(\lambda) = \left( \frac{R_{i}(\lambda)}{R_*} \right)^2 - \left( \frac{R_{\mathrm{cont}}(\lambda)}{R_*} \right)^2 \ ,
    \label{eq:template}
\end{equation}
where $R_{i}(\lambda)$ is the wavelength-dependent transmission spectrum for a single line-absorption species, $i$, and background continuum absorbers, $R_{\mathrm{cont}}(\lambda)$ is the transit radius of just the continuum-forming species, and $R_*$ is the stellar radius. For $R_*$ we adopted a value of $1 \, R_\odot$ in this study. These templates thus describe the contributions of the absorption lines of a species to the transit depth above the spectral continuum. To remove small-scale numerical noise and to create a proper continuum normalisation, values of $d_{\mathrm{t},i}(\lambda)$ smaller than $10^{-7}$ are set to 0.

Figure \ref{fig:spectra_examples1} shows an example of a template spectrum for \ch{Ca+} at a temperature of 4000 K (upper panel, blue line) and the corresponding spectrum of the continuum (red line). The latter is dominated by the strong \ch{H-} continuum, as expected at these high temperatures. By subtracting both spectra, the continuum can be effectively removed, such that the resulting template just provides the transit depth of the \ch{Ca+} lines above the continuum (lower panel).

\subsection{Line masks and subtracted templates}
\label{sec:binary_masks}

The spectral templates discussed in the previous subsection contain spectral lines that can be significantly broadened in order to match or exceed the sampling width of the real-world spectra to which the template is applied. This line broadening in the template, however, also results in an additional source of broadening of the resulting cross-correlation profile or average spectral line. 

This problem can in principle be overcome by using line masks. Rather than using the entire line profile for the cross-correlation procedure, only the position of line centres and a weight for each line are taken into account. A line mask where the same weight for all lines is used is usually referred to as a binary line mask.

Because the line mask is defined only by a list of discrete lines centres, it offers increased freedom in weighing, including, or rejecting individual lines compared to a full spectral template. For example, using masks enables the targeted rejection of specific lines associated with stellar activity, thereby increasing the precision of radial velocity measurements of stars \citep{Dumusque2018}.

Therefore, in addition to the single-species templates, we also provide corresponding line masks. We derived these mask data from the calculated spectra in two ways.

First, we applied a peak-finding algorithm to all continuum-subtracted single-species templates, to record the peak heights of individual lines or line blends. Indistinguishable blends are common due to hyper-fine splitting, meaning that often there is no one-to-one correspondence between the spectral lines present in the transmission spectrum and the underlying line database, necessitating the use of a peak-finding algorithm a posteriori.

To derive effective line positions and depths, we first scanned each single-species template for line peak centres using the continuum-corrected templates (see the lower panel of Fig. \ref{fig:spectra_examples1} for an example) and \texttt{scipy}'s peak-finding algorithm \texttt{find\_peaks}. The configuration parameter for peak prominence is set to 4, meaning that only line peaks that rise 4 km above the local continuum are taken into account -- which is a small fraction of one atmospheric scale height. This removes small false positives caused by numerical noise. This procedure thus also ignores very weak lines that are not strong enough to rise significantly above the continuum at a given temperature. As a result, our line mask data will not contain all possible lines from a given line list, but only lines that are expected to be strong given this atmospheric model. Furthermore, we correlated the detected peak locations with theoretical positions from the original line lists to avoid false positives, recognising that closely separated lines are not individually resolved.

Secondly, we recognise that many spectral lines may overlap with absorption lines of other species, effectively masking their contribution and reducing their prominence relative to other, isolated lines. This negatively affects the result of the cross-correlation analysis because a template that treats a single species as an isolated absorber will overestimate the relative importance of some lines that are masked in the real planet spectrum.  This effect is particularly important for species that are relatively weak, for which many lines may overlap with stronger lines of strongly absorbing transition metals - especially at bluer wavelengths. To quantify this effect, for each of the five temperatures we calculated the transmission spectrum of the planet assuming that it contains all our selected atoms, ions and molecules, as well as the transmission spectrum that contains all but one of these. We subsequently subtracted the two, leaving the effective contribution of the single absorber. Corresponding to Eq.\,\eqref{eq:template}, we obtain the subtracted transit depth for a species $i$ with
\begin{equation}
    d_{\mathrm{s},i}(\lambda) = \left( \frac{R_{\mathrm{all}}(\lambda)}{R_*} \right)^2 - \left( \frac{R_{\mathrm{rm}, i}(\lambda)}{R_*} \right)^2 \ ,
    \label{eq:template_subtract}
\end{equation}
where $R_{\mathrm{all}}(\lambda)$ is the transmission spectrum including all species from Table \ref{tab:opacity_species} and $R_{\mathrm{rm}, i}(\lambda)$ the transit radius where species $i$ has been removed. 

\begin{figure}
  \centering
  \resizebox{\hsize}{!}{\includegraphics{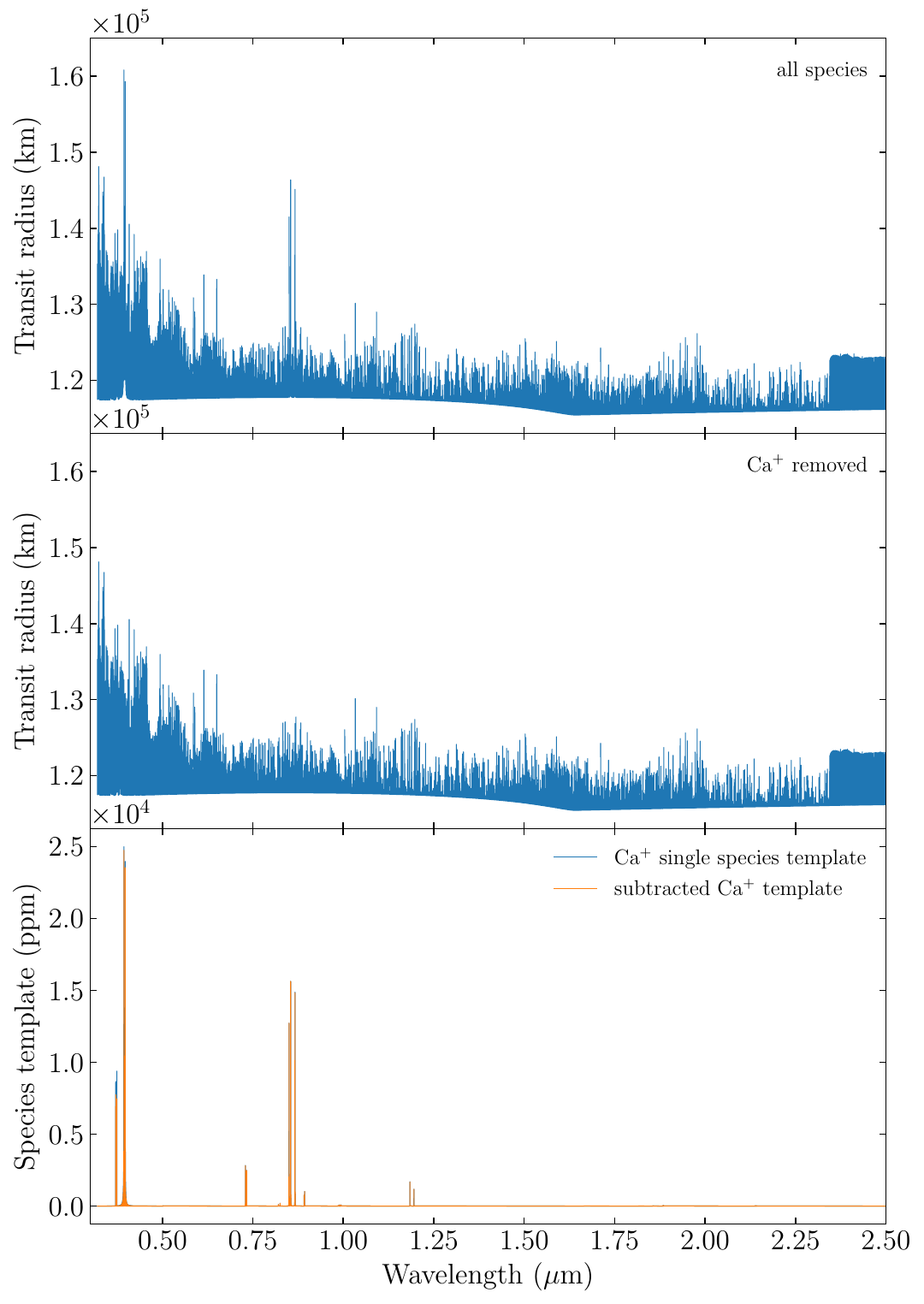}}\\
  \caption{Example of a subtracted template construction of \ch{Ca+} at an atmospheric temperature of $4000 \ \rm{K}$. The upper panel shows the transmission spectrum with all species included ($R_{\mathrm{all}}$), while in the middle panel \ch{Ca+} has been removed as an opacity species ($R_{\mathrm{rm}, \ch{Ca+}}$). The lower panel depicts the final template, $d_{\mathrm{s}, \ch{Ca+}}$, created by subtracting the transit depths of the spectrum of the middle panel from the top-panel one (orange line). Additionally, the single species template $d_{\mathrm{t}, \ch{Ca+}}$ from Fig. \ref{fig:spectra_examples1} is shown for comparison (blue line).}
  \label{fig:spectra_examples2}
\end{figure}

Figure \ref{fig:spectra_examples2} shows an example for a temperature of 4000 K for a spectrum $R_{\mathrm{all}}(\lambda)$ containing all species (top panel) and one where \ch{Ca+} has been removed ($R_{\mathrm{rm}, \ch{Ca+}}(\lambda)$, middle panel). The difference between the transit depths of these contains the missing lines of \ch{Ca+}, from which relative depths are subsequently derived (lower panel).

\begin{figure}
  \centering
  \resizebox{\hsize}{!}{\includegraphics{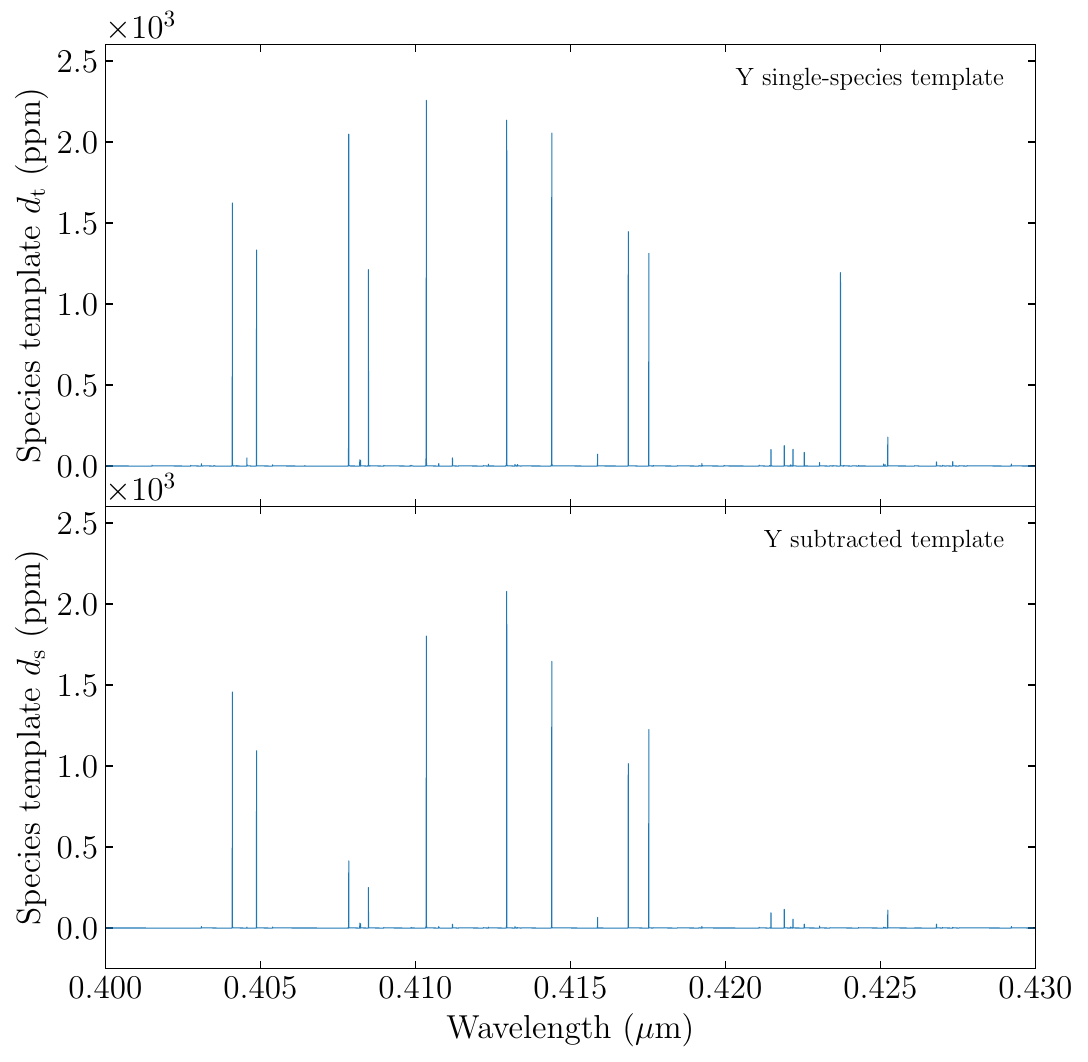}}
  \caption{Example of a normal template and a subtracted spectrum for yttrium at an atmospheric temperature of $4000 \ \rm{K}$. The upper panel shows a small wavelength region of the transmission spectrum for a single-species template of \ch{Y}, i.e. it includes the corresponding continuum-forming species and \ch{Y}. For this template, the underlying continuum has been removed by subtracting the continuum-only transmission spectrum. The lower panel depicts a subtraction of a spectrum with all species with one that contains all species except \ch{Y}. This spectrum essentially highlights the effective contribution of \ch{Y} to the full spectrum.}
  \label{fig:subtract_example}
\end{figure}

The effect of masking by other species is not immediately evident from the bottom panel of Fig. \ref{fig:spectra_examples2}, but is demonstrated in Fig. \ref{fig:subtract_example} where we compare the single-species template spectrum for yttrium with that of the subtracted spectrum. In the subtracted spectrum, some of the lines from the single-species template are entirely missing, while others are weighted differently relative to each other. The missing lines are effectively blocked by the lines of other species. Lines that are still visible but have a considerably smaller height are usually blended with the line wings of other species and are, therefore, only partially visible in the full spectrum.

Finally, we correlated the detected line peak positions from the single-species template with those in the subtracted template spectrum. This allows us to quantify if these lines would actually be visible in a spectrum that contains all species or if the line is partially or totally covered by other species' lines.

\begin{figure}
  \resizebox{\hsize}{!}{\includegraphics{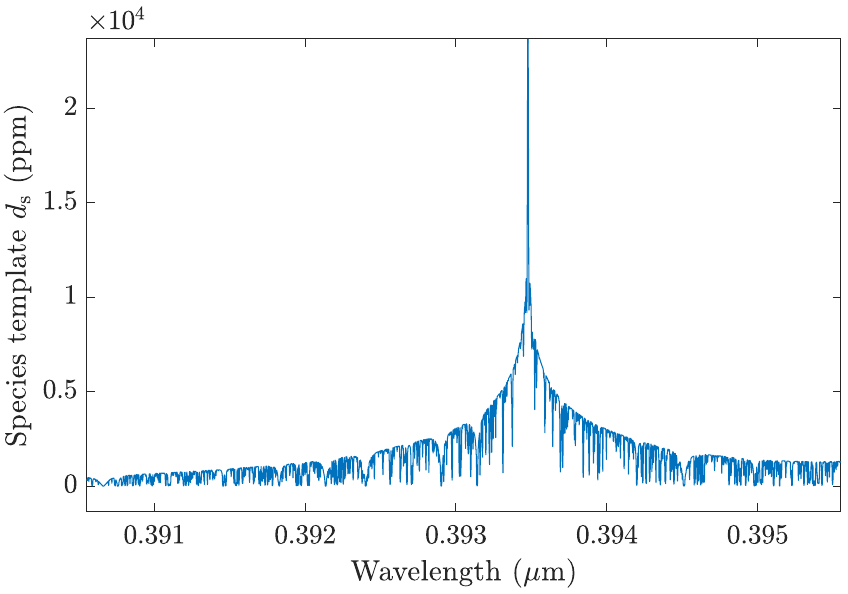}}
  \caption{Example of a subtracted spectrum for \ch{Ca+} at an atmospheric temperature of $4000 \ \rm{K}$. The plot shows a magnification of one of the strong \ch{Ca+} lines, where the line wing has negative imprints of other species' lines.}
  \label{fig:subtract_example2}
\end{figure}

While initially we aimed to use these subtracted spectra as new templates, this proved to be difficult. The subtraction is not perfect because a species' opacity does not linearly transform into the transit radius. Simply subtracting two different transmission spectra therefore does not take the more complex relationship between the transit radius and the absorption coefficients into account, creating small distortions in the line-wings, with negative imprints of absorption by the other species' at a level much smaller than typical line cores. We show such an example for \ch{Ca+} in Fig \ref{fig:subtract_example2}. The negative imprints of other species can clearly be seen in the line wing of the strong \ch{Ca+} line. Therefore, we only used the subtracted spectra to determine the relative depths of line cores, which are used to create line masks, and not full cross-correlation templates.

\section{Results: Templates and masks}

\subsection{Single-species templates}

\begin{figure*}
  \centering
  \includegraphics[width=\hsize]{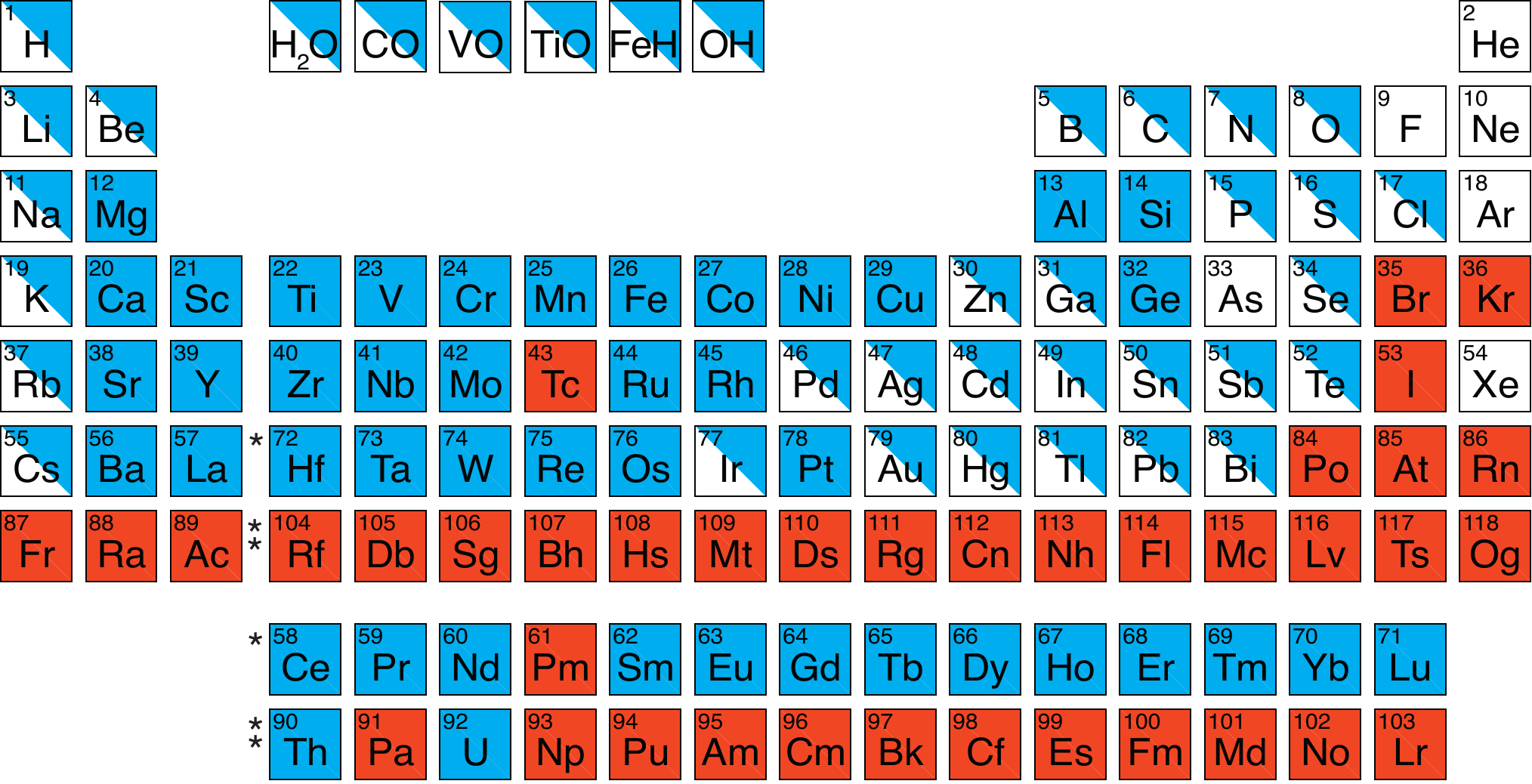}
  \caption{Overview of available templates and line masks. Species fully coloured in blue have templates for both atoms and ions, while for those coloured half in blue, template data for only the atomic form are available. Elements fully coloured white indicate species where line data are available but no absorption lines are present in the wavelength range of the template spectra, whereas species coloured in red are cases where either no line list or chemical data are available. 
  }
  \label{fig:templates_species}
\end{figure*}

Figure \ref{fig:templates_species} shows an overview of all the available templates. Species for which insufficient input data (line lists, thermochemical data) is available are marked in red, while those for which data were available, but that have no detectable absorption lines according to our model at any of the five temperatures, are coloured in white. Species with one or more templates with significant line absorption are denoted in blue, where a species that is only half-coloured has significant line absorption for its neutral, but not for its ion. Plots of all species' templates at three different temperatures are provided in Fig. \ref{fig:templates_all} of the appendix.

All species' cross-correlation templates are available in the CDS data archive in the form of single FITS files. A detailed description of these files can be found in Appendix \ref{sec:appendix_templates}. Besides the templates, we also provide the transmission spectra of the continuum species. This allows the templates $d_{\mathrm{t},i}(\lambda)$ to be easily converted back to the original transmission spectra they are based on.

As mentioned in Sect. \ref{sec:spectra_calculations}, our templates were calculated on a wavenumber grid with a constant step size of 0.01 cm$^{-1}$. This leads to the spectral resolution not being constant over the entire spectral range. For cases, where a template at a constant resolution is required, we therefore provide a Python script that allows the templates to be resampled to a constant-resolution wavelength grid with a user-defined spectral resolution (see Sect. \ref{sec:appendix_resample} for details).

The single species templates were calculated using the Sun's radius as a reference. However, the provided transit depths above the continuum given by Eq. \eqref{eq:template} can easily be converted to any other stellar radius $R_*$ by simply multiplying them with a constant factor $\left(R_\odot/R_*\right)^2$.

\subsection{Line masks}

For the line masks, we recorded the locations and line depths of both the single-species template as well as the subtracted spectra (see Sect. \ref{sec:binary_masks}). Because these values are based on the continuum-corrected templates, these values should be understood as the transit depths above the underlying spectral continuum. 

The given transit depths for all lines can be used as weights for cross-correlation calculations. They also allow a subset of specific lines to be easily selected, for example, if the cross-correlation should be focused on a sample that only includes the strongest lines.

\begin{figure}
  \resizebox{\hsize}{!}{\includegraphics{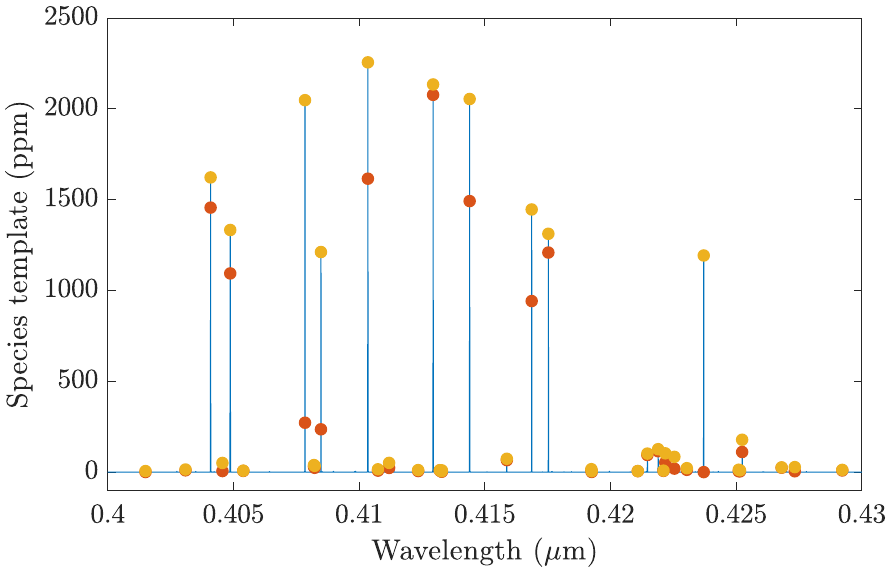}}
  \caption{Example of a line mask of \ch{Y} at an atmospheric temperature of 4000 K. The blue line represents the single-species template from Fig. \ref{fig:subtract_example}. Orange circles denote the detected line peaks and the corresponding line heights. The red circles represent the line peaks in the subtracted template. As can be noticed, many lines are significantly down-weighted or even totally absent due to being covered by other species.}
  \label{fig:binary_example}
\end{figure}

Figure \ref{fig:binary_example} shows an example of the line mask for \ch{Y} at a temperature of 4000 K. The orange circles represent the detected line peak positions and heights in the singe-species template, while the red circles denote the corresponding data based on the subtracted template data. The figure demonstrates that many lines of \ch{Y} in this wavelength range are significantly reduced or even invisible in the subtracted template. These lines are effectively masked by other species and their contribution in a cross-correlation procedure should, therefore, be decreased.

The line masks are saved as simple ASCII tables. They can be found in the CDS data archive, and their contents are described in detail in Appendix \ref{sec:appendix_line_masks}. Line masks are only provided for the atoms and ions. Due to very high number of spectral lines for the molecules \ch{CO}, \ch{H2O}, TiO, VO, FeH, and OH, many of which are blended, identifying line centre positions is challenging. Therefore, we decided not to provide line masks for these species.

The subtracted line masks can be used when there is suspicion of significant blending. They take into account the effects of micro-physical broadening of lines (e.g. thermal or pressure broadening). Other sources of broadening, for example due to limited resolving power, may aggravate blending, but in such cases the absorption of individual lines integrates.

To apply line masks or subtracted line masks to real data as part of a cross-correlation analysis, the tabulated weight for each line position should be used to evaluate the cross-correlation. This requires that the weight of each line be distributed over the two nearest spectroscopic pixels, depending on how far the centre of the line is from the centres of each of these (equivalent to linear interpolation). In the case of lower resolving power, this implicitly acts to integrate the absorption strength of multiple blended lines. For a typical template of an atomic species, the weight of most spectral pixels will remain zero, meaning that the computation of the cross-correlation can be significantly faster than when using a spectral template. Besides providing an increase in effective spectral resolving power of the cross-correlation signal, this is a key advantage of using like masks.

\section{Application to WASP-121 b data}
\label{sec:xcrorr_application}

\begin{figure*}
  \resizebox{1.0\hsize}{!}{\includegraphics[trim=2.5cm 0.0cm 2.5cm 0.0cm,clip]{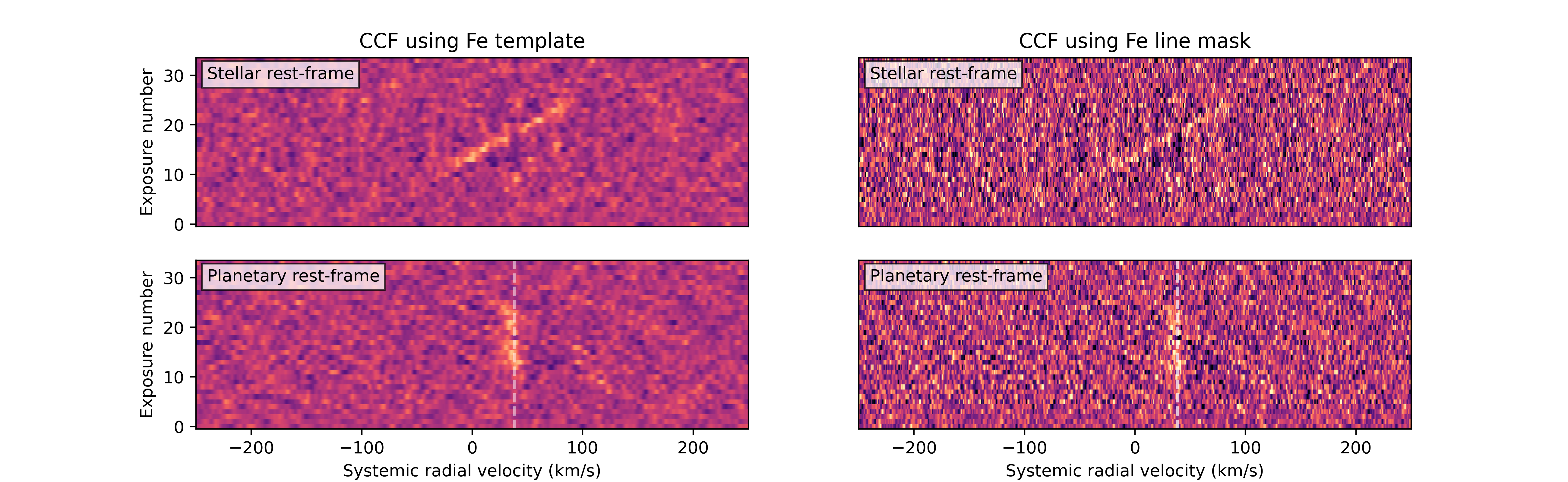}}\\
  \caption{Cross-correlation diagrams showing Fe absorption in the transmission spectrum of WASP-121 b, data published previously in \citet{Hoeijmakers2020}. The cross-correlation function is evaluated using either a spectral template assuming an isothermal atmosphere of 2500~K (left panels) or a line mask (right panels), derived from the template. The cross-correlation function is sampled in steps of 1 km s$^{-1}$ in systemic radial velocity for each exposure in the time series, and the cross-correlation functions of the second and third night of observations were interpolated onto the phase range of the first night, allowing them to be co-added to increase the signal-to-noise. The cross-correlation functions in the top panels are left in the stellar rest frame, while those in the bottom panels are shifted to the planetary rest frame (assuming an orbital velocity of 221 km $\mathrm s^{-1}$). The dashed vertical line indicates the systemic radial velocity, which is equivalent to the rest-frame velocity of the planet.}
  \label{fig:Compare2Dccfs}
\end{figure*}

\begin{figure}
  \resizebox{1.0\hsize}{!}{\includegraphics{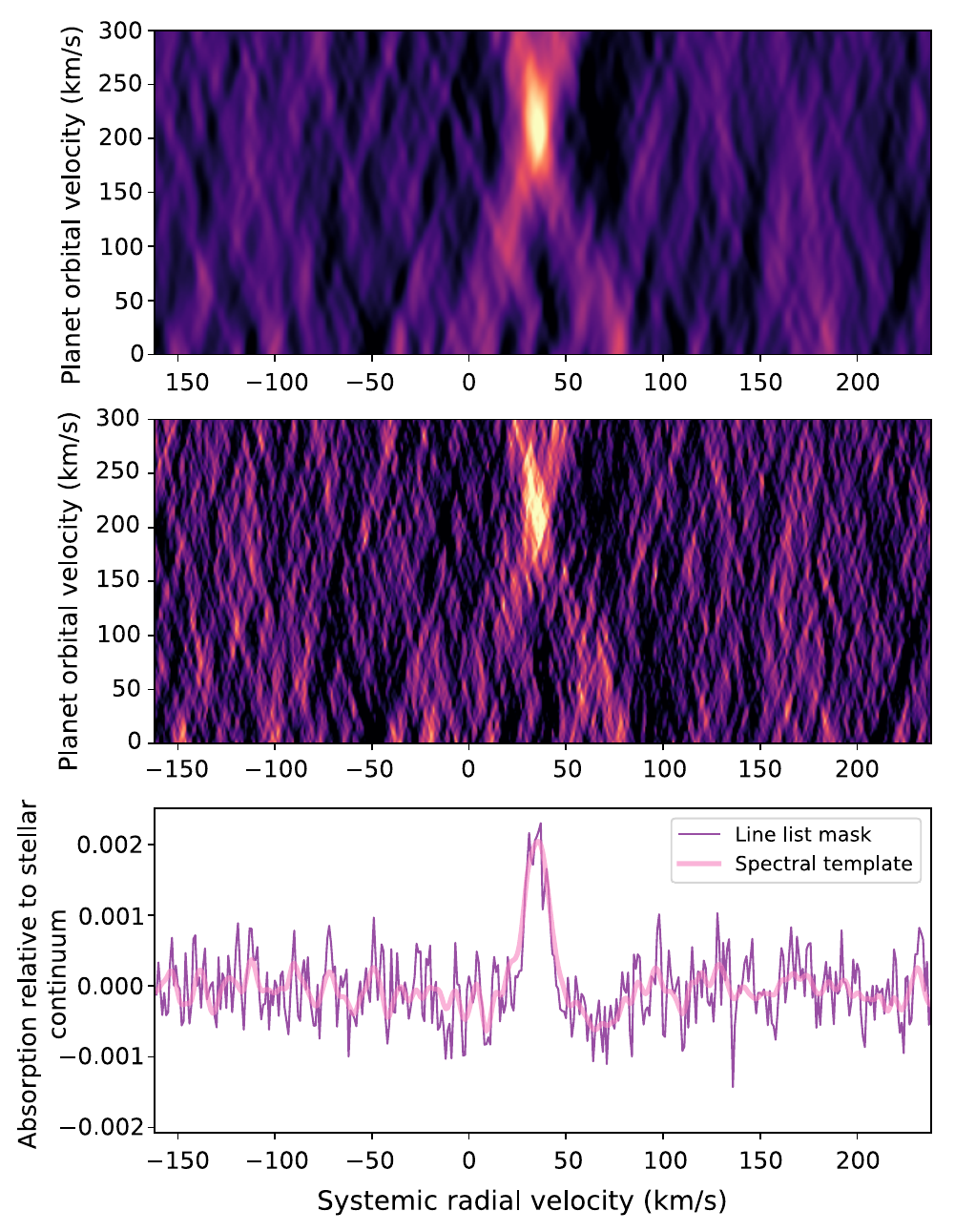}}\\
  \caption{Cross-correlation diagrams integrated over a range of orbital velocities, producing velocity-velocity maps typically shown in similar studies based on cross-correlation spectroscopy. These maps were derived using either the spectral template (top panel) or the line mask (middle panel). The cross-correlation function is sampled in steps of 1 km s$^{-1}$ in systemic radial velocity and orbital velocity. The cross-correlation functions integrated in the rest frame of the planet (assuming an orbital velocity of 221 km $\mathrm s^{-1}$, bottom panel) demonstrate that the line mask approach results in an equivalent, but higher-resolution, measurement of the Fe absorption line.}
  \label{fig:kpvsys}
\end{figure}

We tested and compared the performance of these templates and line-masks on real high-resolution spectra of WASP-121\,b, which has estimated values for radius, mass, and zero-albedo equilibrium temperature of $1.807 \pm 0.039 \, R_\mathrm{J}$, $1.183\pm 0.063 \, M_\mathrm{J}$, and $2358 \pm 52$ K. \citet{Hoeijmakers2020} detected Fe\,I in the transmission spectrum of this ultra-hot Jupiter using observations obtained with the HARPS spectrograph at ESO's 3.6 m telescope during three transit events. We replicated this analysis, using the same pipeline-reduced data, as well as the same analysis steps. The data analysis includes normalisation of the spectra in the time series, outlier rejection, removal of telluric absorption lines with \textsc{Molecfit} \citep{Smette2017sgvi.confE..41S}, correction of time-dependent low-frequency variations (colour), and correction of barycentric and Keplerian velocity variations. We refer the reader to \citet{Hoeijmakers2019A&A...627A.165H} and \citet{Hoeijmakers2020} for detailed descriptions of these steps \footnote{The analysis of \citet{Hoeijmakers2020} is nearly identical to the analysis presented here. The main differences are the usage of a hotter (2500 K) template that better matches the temperature at the terminator \citep{Evans2022} and the repeating of a manual removal of regions with systematic noise, leading to a different number of rejected spectral pixels. This has no significant consequence for the present comparison between the template and the line mask of Fe.}.  Figure \ref{fig:Compare2Dccfs} shows resulting velocity-velocity maps of the cross-correlation function as a function of the orbital velocity, $K_{\rm{p}}$, and the systemic velocity, $v_{\rm{sys}}$. We compare the performance of the spectral template and the derived line mask of Fe\,I, which are in excellent agreement. Compared to the signal presented by \citet{Hoeijmakers2020}, the Fe\,I line in Fig. \ref{fig:kpvsys} is approximately 50\% less deep. We verified this happens because the 2500 K template used here assigns more weight to weaker lines that are prominent at higher temperatures, compared to the 2000 K template used by \citet{Hoeijmakers2020}.

\section{The effect of surface gravity}
\label{sec:gravity}

\begin{figure}
  \resizebox{1.0\hsize}{!}{\includegraphics{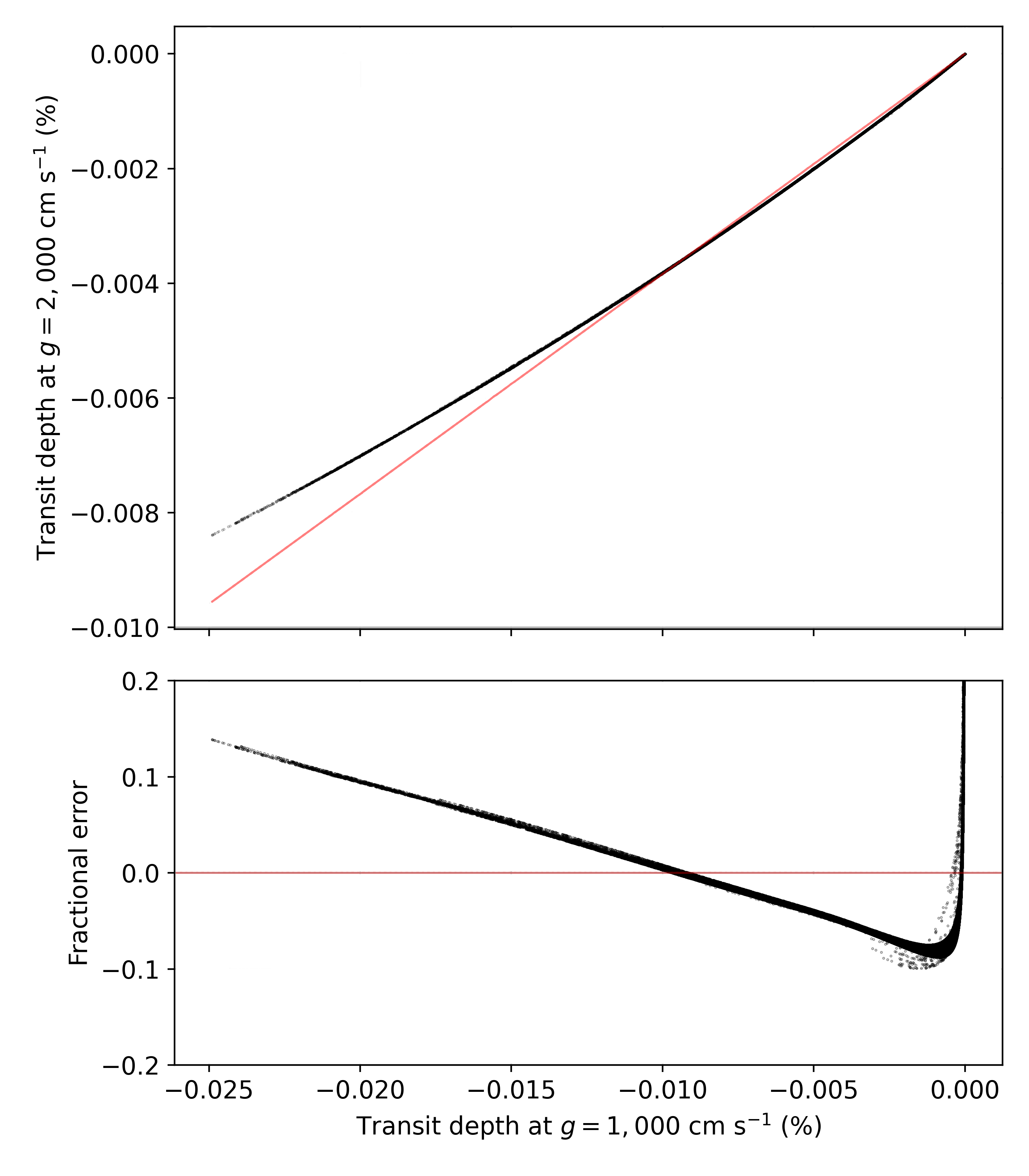}}\\
  \caption{Template of neutral iron computed at two different values for the surface gravity, $g$, plotted against itself (top panel) and their relative differences (bottom panel). To first order, a change in gravity can be represented by a constant scaling (dashed line, linear fit). Departures of up to 15\% occur for the weakest and strongest lines, which span a multitude of pressure scale heights. However, we find that they cause differences in the strength of the cross-correlation of no more than 0.5\%.}
  \label{fig:gravity_test}
\end{figure}

To compute these templates and masks, the surface gravity was kept fixed at $g=2000$\,cm\,s$^{-2}$, a factor of two higher than some of the most inflated hot Jupiters. To first order, the transmission spectrum depends linearly on the surface gravity via the scale height \citep[see e.g.][]{HK2017}. The cross-correlation, being invariant to the scaling of the template, is therefore not expected to be strongly sensitive to the choice of surface gravity. However, departures from simple scaling may be significant if the template contains lines with depths that span large ranges of pressure (see Fig. \ref{fig:gravity_test}). 

We tested that the choice of a fixed value of the surface gravity does not impair the application of these templates by injecting templates with different values of the surface gravity into a time series of high-resolution transit observations of the hot Jupiter WASP-49\,b, data originally published by \citet{Wyttenbach2015}. We injected models with $g = 1000$ and $g = 2000$\,cm\,s$^{-2}$, respectively, and repeated the cross-correlation analysis. We measure the signal-to-noise ratio of the injected cross-correlation as the peak amplitude of the difference between the injected and non-injected cross-correlation function divided by the noise in the cross-correlation function at velocities away from the injected signal. We observe that the signal-to-noise ratio decreases by no more than 0.5\% when correlating with either template compared to the other. Due to the scale-invariance of the cross-correlation and the weak dependence on signal-to-noise on 10\%-level errors in the relative line depths, we conclude that this grid of templates with a constant surface gravity can be applied with negligible loss of sensitivity for a wide range of surface gravities, including highly inflated planets such as WASP-121\,b or more compact planets such as WASP-18\,b.

\section{Summary}

In this work we have presented a grid of cross-correlation templates for a large variety of atoms and ions, as well as the molecules water, carbon monoxide, vanadium oxide, titanium oxide, iron hydride, and the hydroxyl radical. These templates were designed to perform cross-correlation calculations of observational data of ultra-hot Jupiters. They in principle allow the template species to be detected in exoplanet spectra, if they are present and if the signal-to-noise ratio of the measured spectra is high enough. By using standard templates, the cross-correlation results of different observations can be compared in a systematic and quantitative way. 

We provide the templates for five different temperatures in total, from 2000 K to 5000 K, and for solar element abundances. The provided templates are aimed at the detection of chemical species, rather than retrieving their abundances or other atmospheric properties, such as temperatures. 

In addition to the usual templates, we also created line masks. These masks consist of line positions and relative depths of absorption lines, but do not contain information on the actual line profiles. We also included special versions of these line masks for each species that take potential blending effects with other species into account. This blending can reduce the relative importance of certain lines that are affected by other species' spectral lines and, thus, should not be given their full weight in a cross-correlation procedure. 

By applying the templates to existing observational data, we show that they can detect atomic iron in the atmosphere of WASP-121 b, which has previously been found by \citet{Hoeijmakers2020}. Since the templates and line masks have been calculated for a specific surface gravity of $g = 2000$\,cm\,s$^{-2}$, we tested the impact for cases where $g$ differs from that value. We find that within the expected surface gravity range of ultra-hot Jupiters, the cross-correlation signal differs by no more than 0.5\% compared to templates calculated with the correct $g$. All templates and line masks are publicly available on the CDS data archive server.

\begin{acknowledgements}
  D.K. and S.L.G. acknowledge financial support from the Center for Space and Habitability (CSH) of the University of Bern. 
\end{acknowledgements}

\bibliographystyle{aa}
\bibliography{references.bib}

\begin{appendix} 
\section{Data overview and file formats}

\subsection{Cross-correlation templates}
\label{sec:appendix_templates}

As mentioned in Sect. \ref{sec:singe_species_templates}, the templates are provided in the form of FITS files. Each FITS file contains the spectral grid (wavelengths in nm) and the quantity
\begin{equation*}
    -d_{\mathrm{t},i}(\lambda) = \left( \frac{R_{\mathrm{cont}}(\lambda)}{R_\odot} \right)^2 - \left( \frac{R_{i}(\lambda)}{R_\odot} \right)^2  \ ,
    \label{eq:template_fits}
\end{equation*}
which, following Eq. \eqref{eq:template}, is the negative transit depth above the spectral continuum. This convention of treating absorption as a negative contribution is identical to previous approaches by, for example, \citet{Hoeijmakers2019A&A...627A.165H}.

The templates are given for each species and atmospheric temperature individually. Each filename contains the species, the metallicity, and the atmospheric temperature used to generate the template. It has the following naming scheme: \\
\texttt{xxy\_m\_i\_t\_jjjj.fits} \ ,\\
where \texttt{xx} is the chemical symbol of the element (e.g. \ch{Mg} for magnesium), \texttt{y} the ionisation stage (I denotes the neutral and II the singly ionised atom), \texttt{i} the metallicity [M/H]\footnote{In this study we only provide templates for the case of $[M/H]=0$, i.e. solar element abundances.}, and \texttt{jjjj} the isothermal, atmospheric temperature. For the two molecules \ch{CO} and \ch{H2O}, \texttt{xx} refers to their molecular formulas, while the quantity \texttt{y} is not used. 
All templates are hosted on the CDS database. The templates can either be downloaded individually or as a  compressed archive file (\texttt{templates.tar.gz}). 

In Python, these FITS files can be simply read using the \texttt{astropy} package:\\

\noindent \texttt{from astropy.io import fits}\\
\noindent \texttt{import numpy as np}\\

\noindent \texttt{hdul = fits.open(file\_path)}\\
\noindent \texttt{template = hdul[0].data}, \\

\noindent where \texttt{template} is a two-dimensional \texttt{numpy} array. Its first dimension $\texttt{template[0,:]}$ refers to the wavelength in units of nm and the second dimension $\texttt{template[1,:]}$ to the quantity $-d_{\mathrm{t},i}(\lambda)$. 

Figure \ref{fig:templates_all} provides an overview of all templates in the form mentioned above. There, the quantity $-d_{\mathrm{t},i}(\lambda)$ is plotted in units of ppm for three different atmospheric temperatures. Plots without a line denote species where either no line list or thermochemical data are available. 

In addition to the aforementioned templates, we also provide the continuum transmission spectra $R_{\mathrm{cont}}(\lambda)$ for each temperature (see the upper panel of Fig. \ref{fig:spectra_examples1}). Using these continua together with the species templates, the original single-species transmission spectra $R_{i}(\lambda)$ as shown in the upper panel of Fig. \ref{fig:spectra_examples1} can be easily reconstructed.

\subsection{Template resampling}
\label{sec:appendix_resample}

\begin{figure}
  \resizebox{\hsize}{!}{\includegraphics{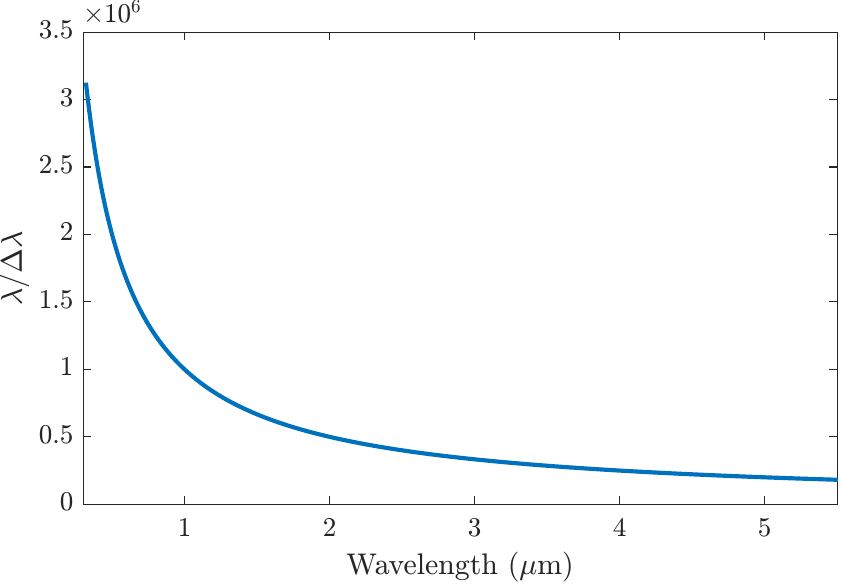}}
  \caption{Spectral resolution as a function of wavelength for a constant step size of  $0.01 \ \rm{cm^{-1}}$ in wavenumber space.}
  \label{fig:spectral_resolution}
\end{figure}

As mentioned in Sect. \ref{sec:spectra_calculations}, our templates are given with a constant wavenumber step size of 0.01 cm$^{-1}$. This implies that the spectral resolution $R=\lambda/\Delta \lambda$ is not constant over the entire tabulated wavelength range from the visible to the infrared (see Fig. \ref{fig:spectral_resolution}).

Since for some applications it might be advantageous to have templates at a fixed spectral resolution, we provide an additional Python script that can resample the templates to a constant-resolution wavelength grid with a user-specified spectral resolution. The Python script also allows to extract only a certain wavelength range
from the original templates in order to make them easier to apply to specific spectrographs. The script can be found in the GitHub repository \url{https://github.com/daniel-kitzmann/XCorrTemplateResample}.

We note, however, that the down-sampling to a lower resolution than offered by our original templates is only done with a simple integration. For a more suitable application to specific instruments, the high-resolution data should be convolved with a proper instrument line profile function to take the characteristics of a certain spectrograph more accurately into account.\\

\subsection{Line masks}
\label{sec:appendix_line_masks}

In addition to the cross-correlation templates, we also provide the line masks, described in Sect. \ref{sec:binary_masks}. These masks are saved as simple ASCII text files. For each species, two different files are provided:\\
\texttt{xxy\_m\_i\_t\_jjjj.dat} \\
\texttt{xxy\_m\_i\_t\_jjjj\_subtract.dat} \ ,\\
where the first one refers to the line masks based on the single-species templates and the second file to the mask derived from the subtracted spectra. The naming scheme is identical to the one of the cross-correlation template files explained in the previous subsection. The line masks are only provided for atoms and ions, but not for the molecules \ch{CO}, \ch{H2O}, VO, TiO, FeH, and OH.

Each file contains a header that also lists the corresponding line list the species' opacity calculations are based on. The data themselves are tabulated in two columns. The first column refers to the wavelength of the line centres in units of nm and the second one the corresponding line weight. As explained in Sect. \ref{sec:binary_masks}, these weights are the transit depths above the continuum of either the single-species templates $d_{\mathrm{t},i}(\lambda)$ or the subtracted ones ($d_{\mathrm{s},i}(\lambda)$).

Like the single-species templates, the line mask files can either be downloaded individually or as a compressed archive file (\texttt{masks.tar.gz}).

\begin{figure*}[b!]
  \centering
  \resizebox{0.90\hsize}{!}{\includegraphics{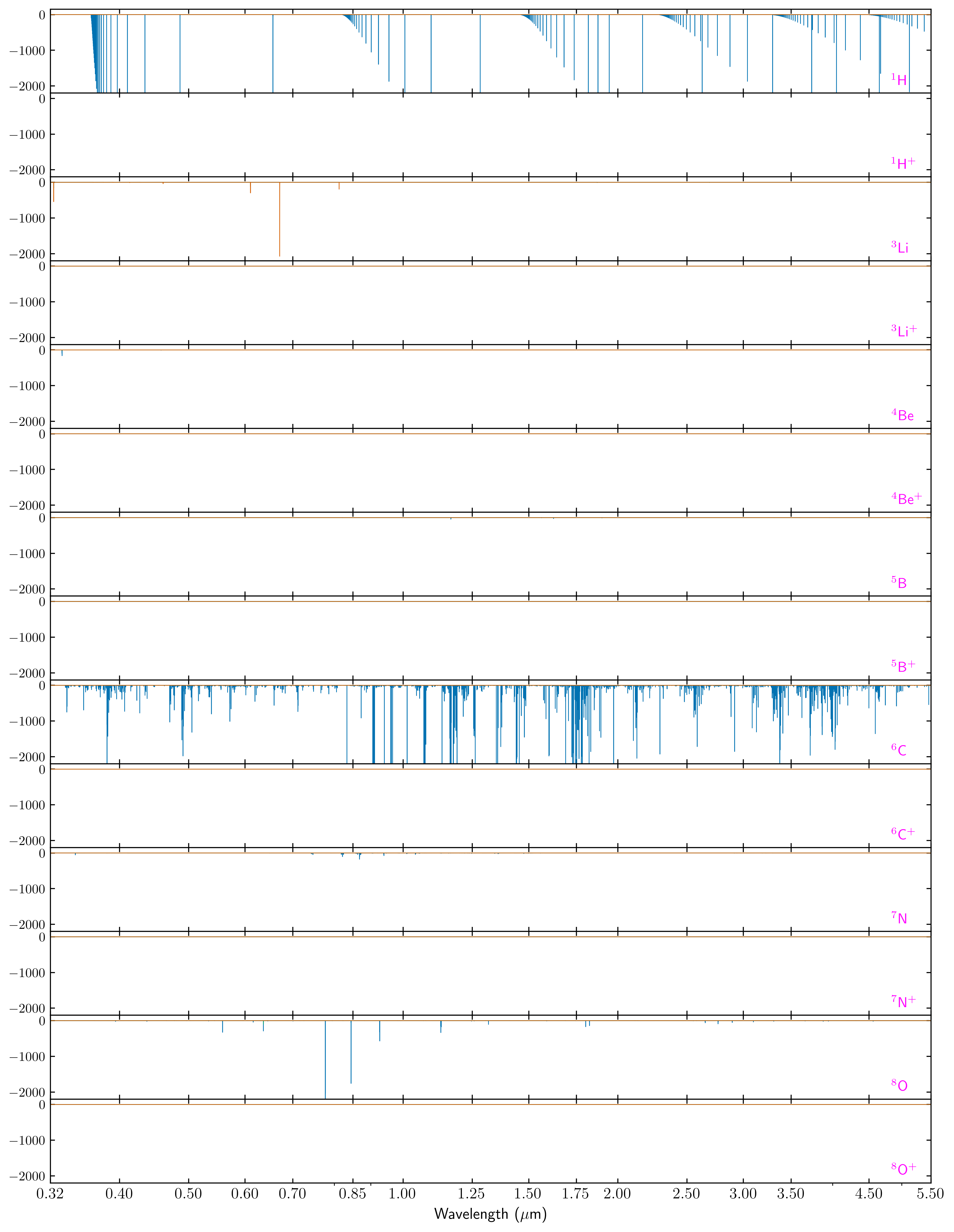}}\\
  \caption{Cross-correlation templates of all atoms, ions, and molecules considered in this study. The plots show negative transit depth {above the spectral continuum} ($-d_{\mathrm{t}}$) of a 1.5 R$_\mathrm{J}$ planet orbiting a 1 R$_\odot$ star in ppm as a function of wavelength. Templates for three different temperatures are depicted: 5000 K (blue), 3000 K (green), and 2000 K (orange). Figures without a line symbolise species where either no line list data or thermochemical data are available.}
  \label{fig:templates_all}
\end{figure*}

\begin{figure*}
  \centering
  \resizebox{0.88\hsize}{!}{\includegraphics{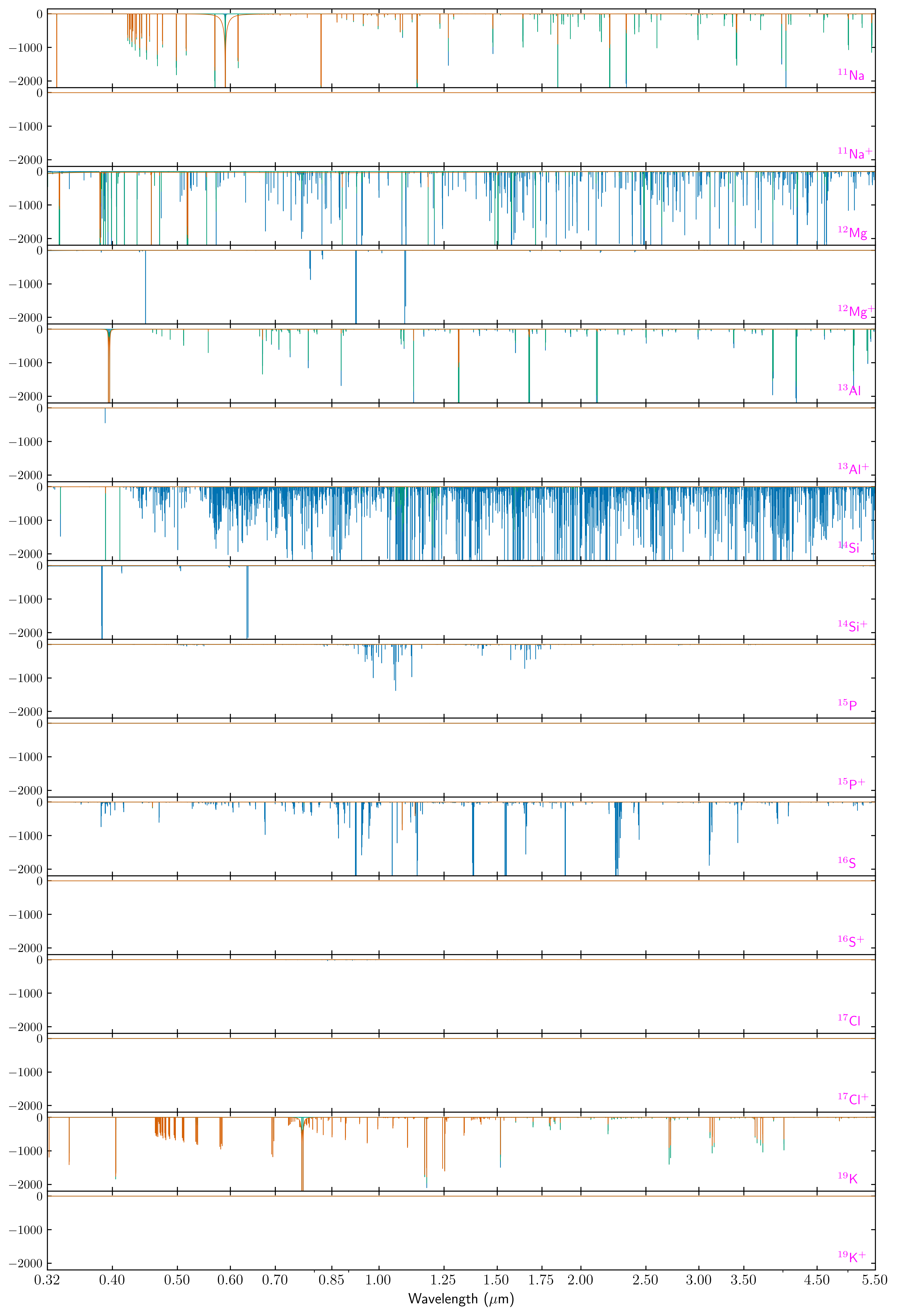}}\\
  \figurenum{\ref{fig:templates_all}}
  \caption{continued.}
\end{figure*}

\begin{figure*}
  \centering
  \resizebox{0.88\hsize}{!}{\includegraphics{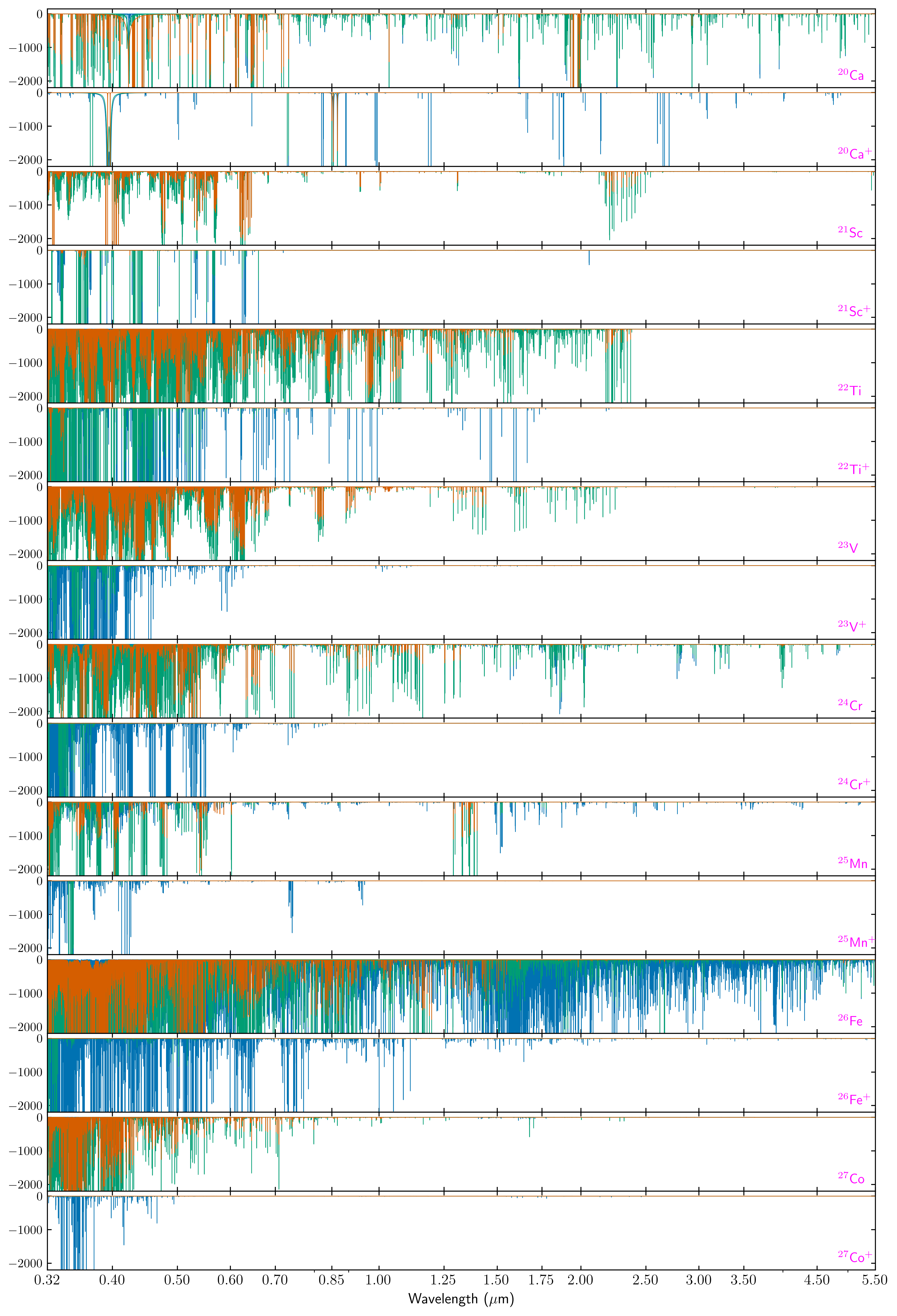}}\\
  \figurenum{\ref{fig:templates_all}}
  \caption{continued.}
\end{figure*}

\begin{figure*}
  \centering
  \resizebox{0.88\hsize}{!}{\includegraphics{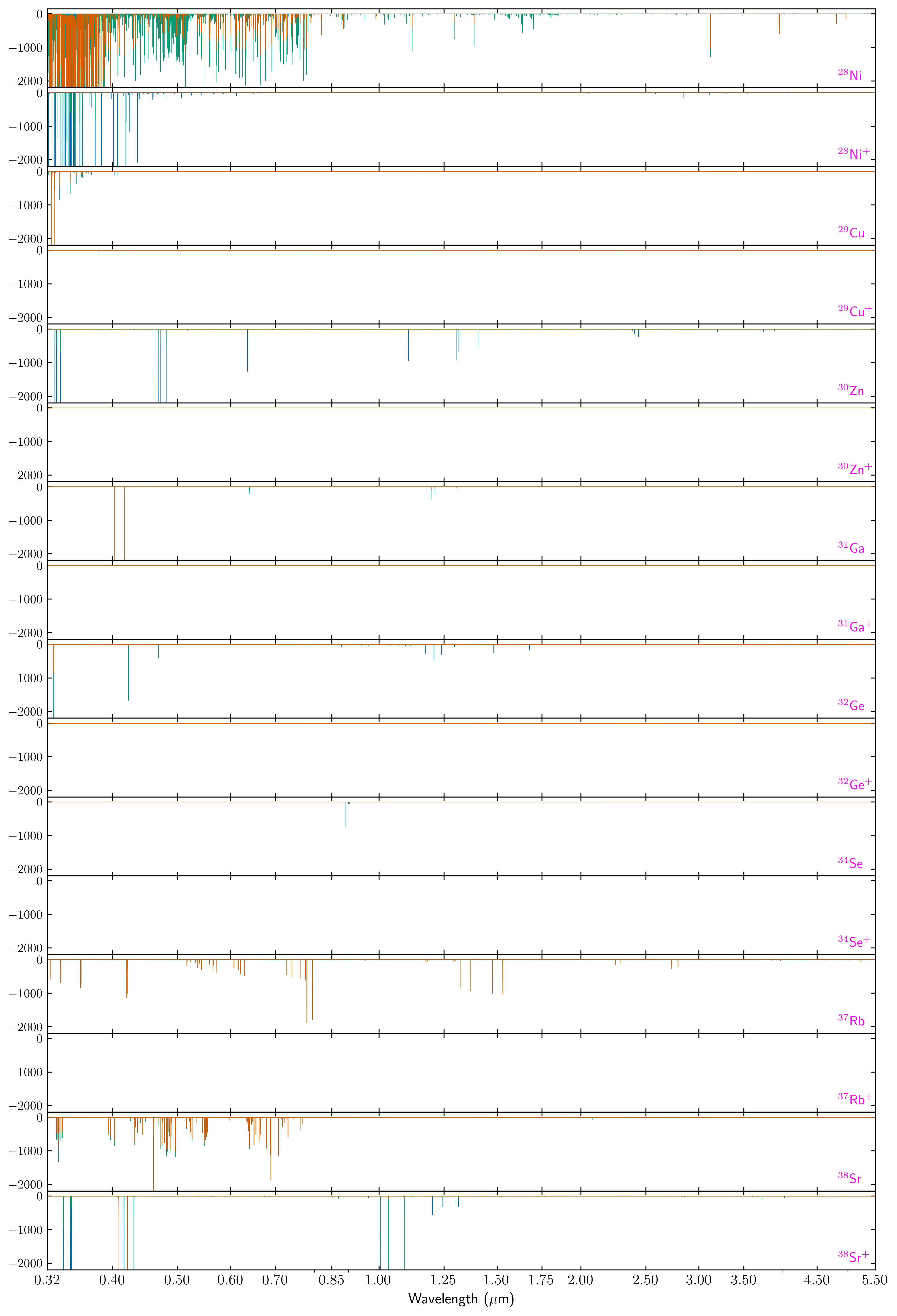}}\\
  \figurenum{\ref{fig:templates_all}}
  \caption{continued.}
\end{figure*}

\begin{figure*}
  \centering
  \resizebox{0.88\hsize}{!}{\includegraphics{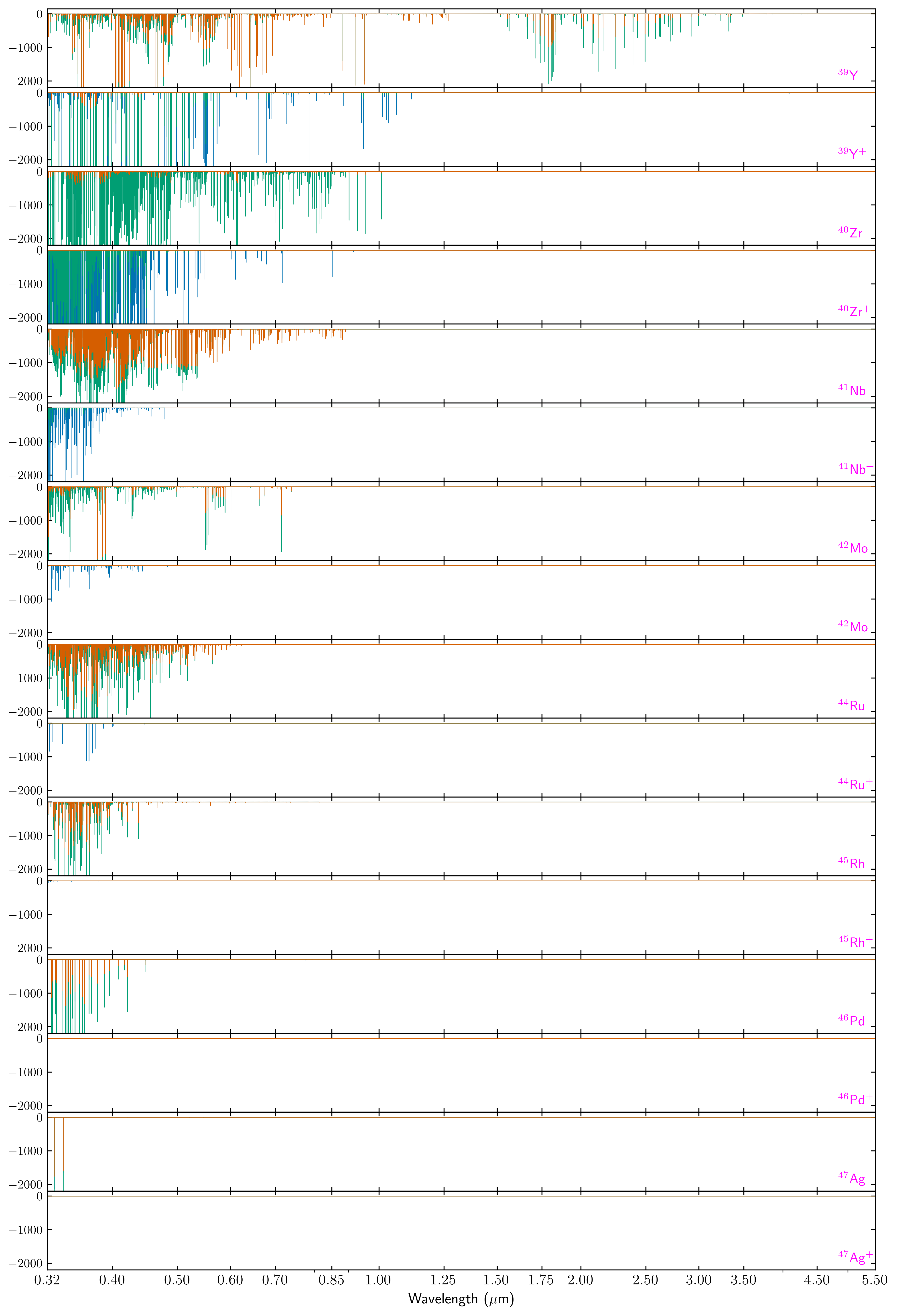}}\\
  \figurenum{\ref{fig:templates_all}}
  \caption{continued.}
\end{figure*}

\begin{figure*}
  \centering
  \resizebox{0.88\hsize}{!}{\includegraphics{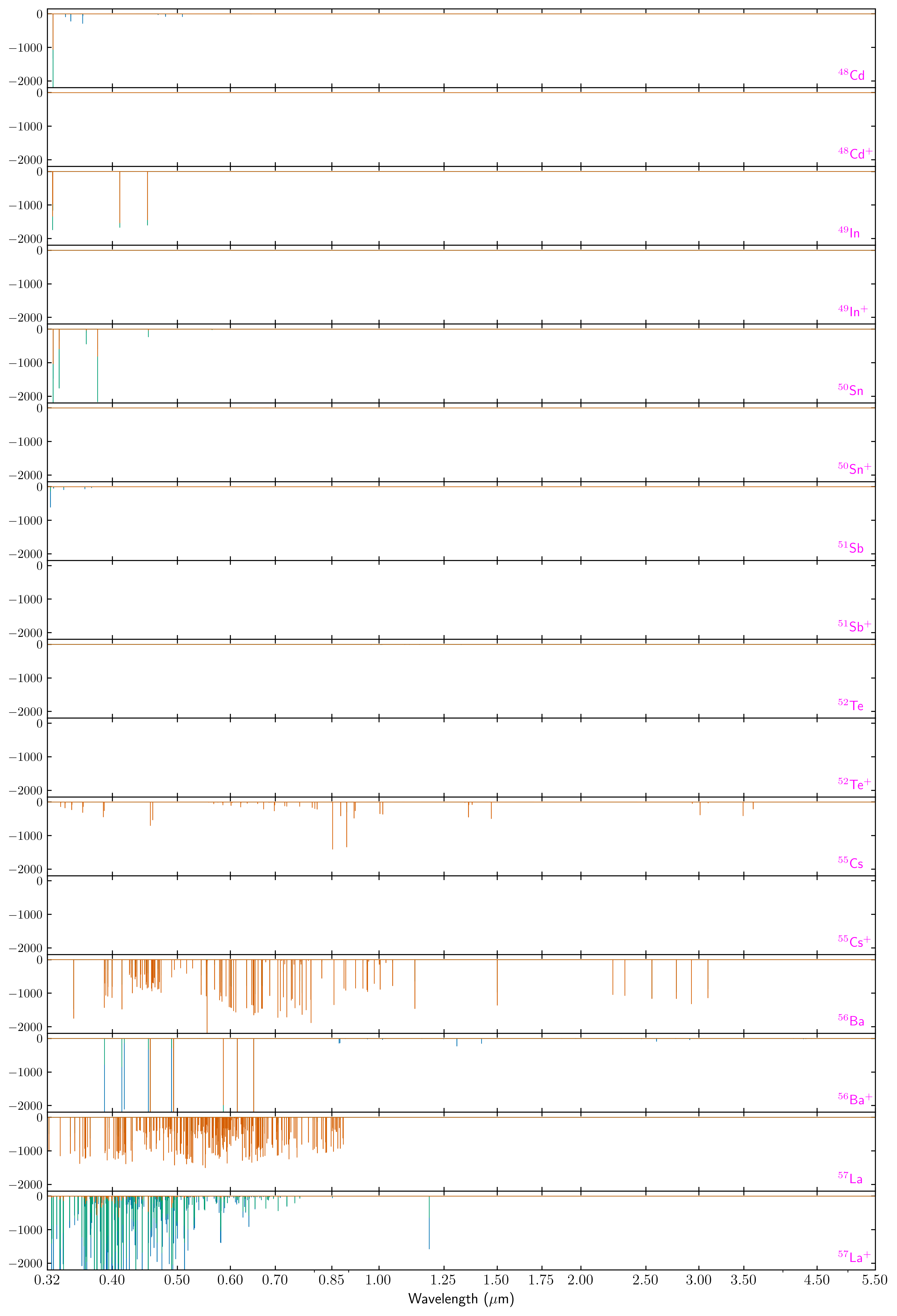}}\\
  \figurenum{\ref{fig:templates_all}}
  \caption{continued.}
\end{figure*}

\begin{figure*}
  \centering
  \resizebox{0.88\hsize}{!}{\includegraphics{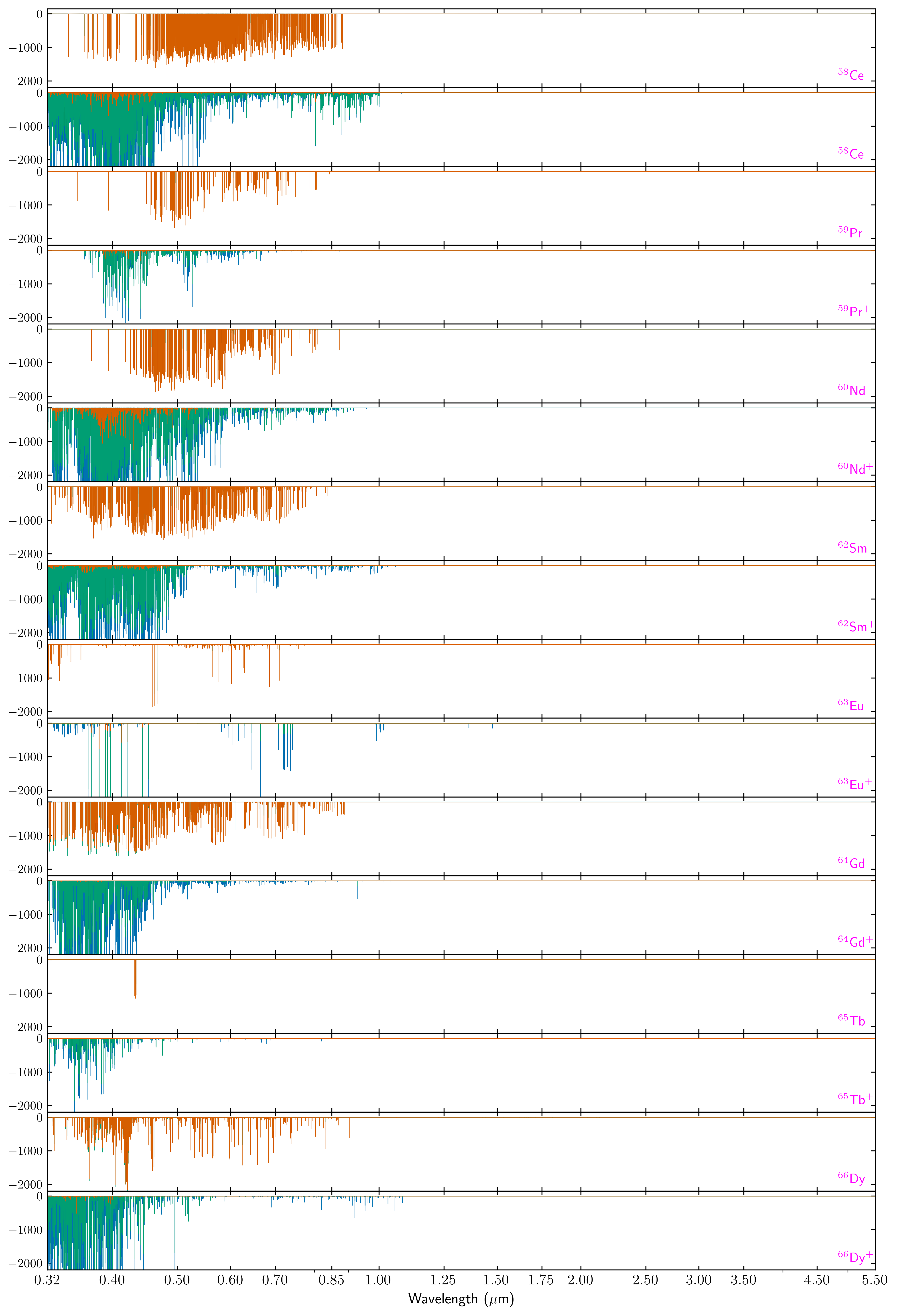}}\\
  \figurenum{\ref{fig:templates_all}}
  \caption{continued.}
\end{figure*}

\begin{figure*}
  \centering
  \resizebox{0.88\hsize}{!}{\includegraphics{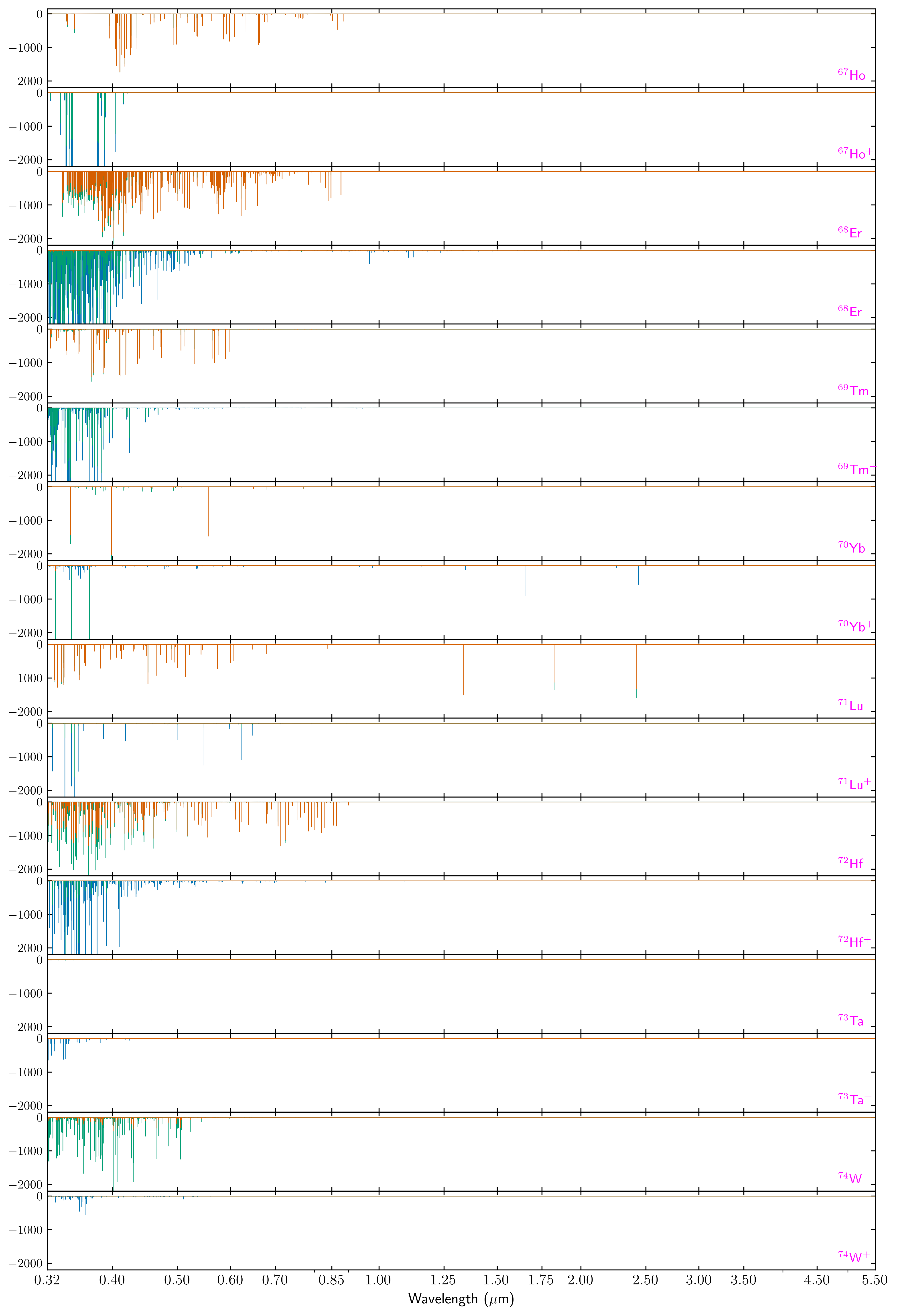}}\\
  \figurenum{\ref{fig:templates_all}}
  \caption{continued.}
\end{figure*}

\begin{figure*}
  \centering
  \resizebox{0.88\hsize}{!}{\includegraphics{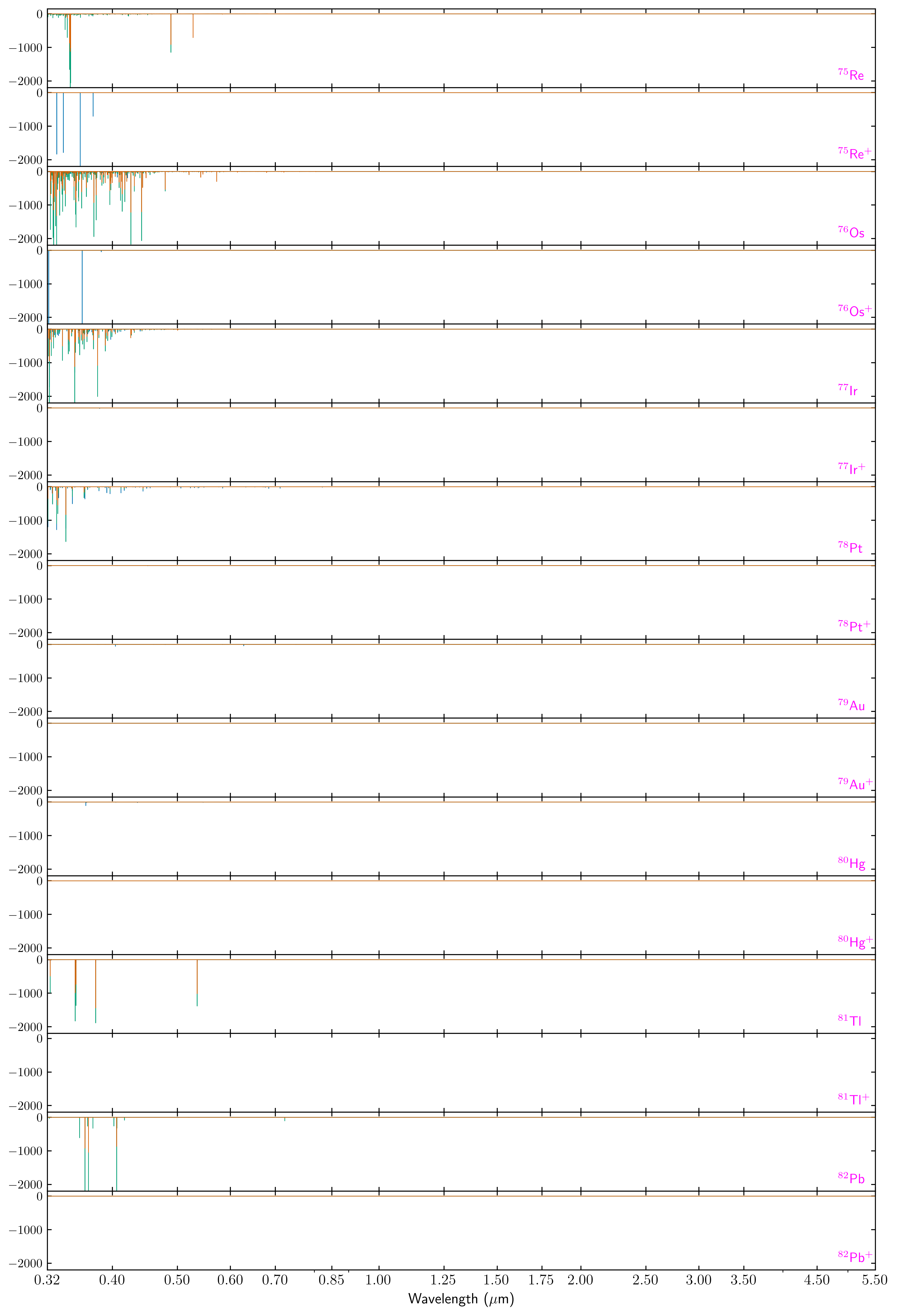}}\\
  \figurenum{\ref{fig:templates_all}}
  \caption{continued.}
\end{figure*}

\begin{figure*}
  \centering
  \resizebox{0.88\hsize}{!}{\includegraphics{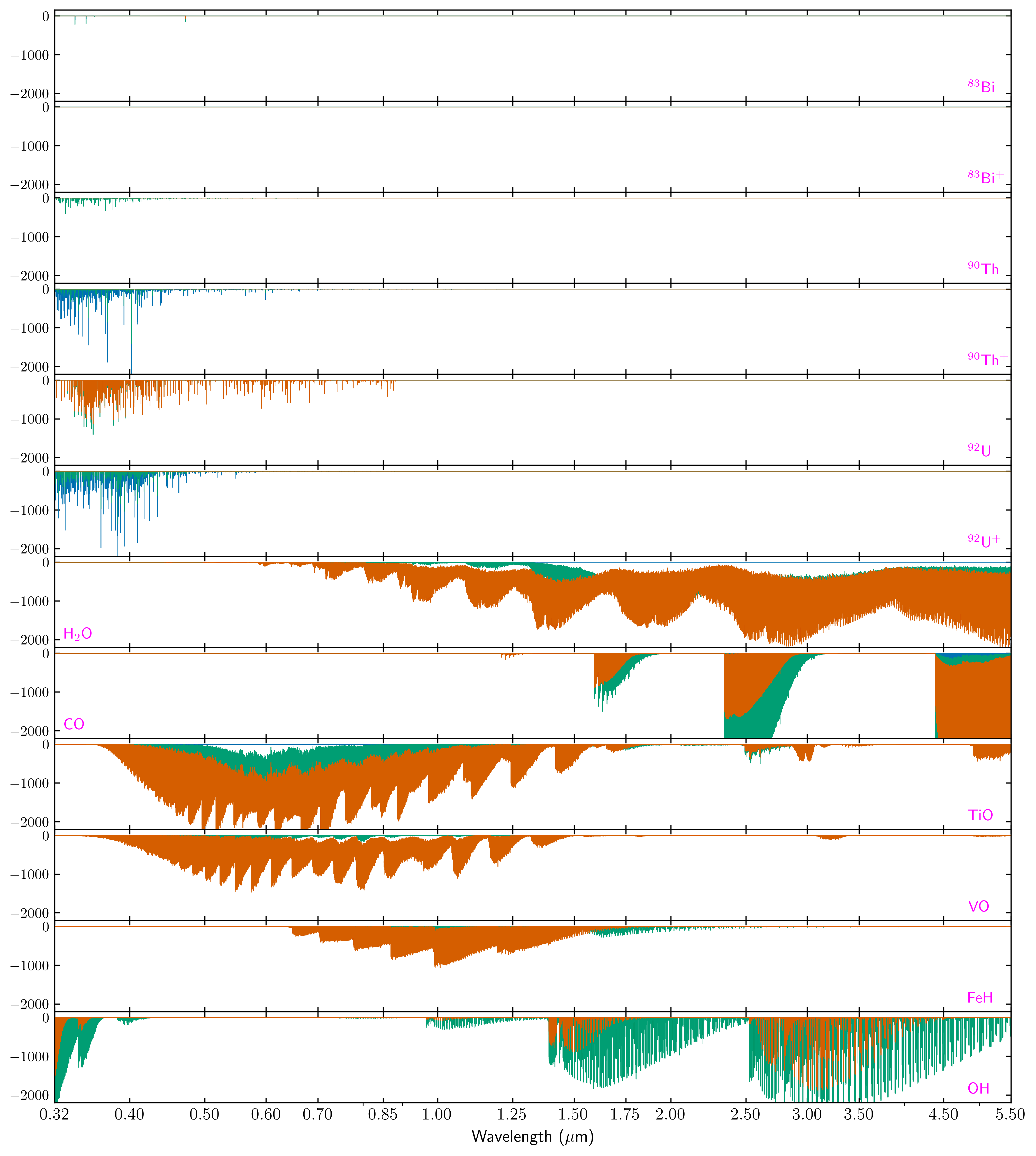}}\\
  \figurenum{\ref{fig:templates_all}}
  \caption{continued.}
\end{figure*}

\end{appendix}

\end{document}